\documentclass[%
reprint,
amsmath,amssymb,
aps,
pra,
longbibliography,
]{revtex4-2}
\usepackage{silence}
\usepackage{graphicx}% Include figure files
\usepackage{dcolumn}% Align table columns on decimal point
\usepackage{bm}% bold math
\usepackage{hyperref}% add hypertext capabilities
\usepackage{tikz}
\usepackage{xcolor}
\usetikzlibrary{arrows.meta, positioning, calc}
\usetikzlibrary{calc}
\usepackage{ragged2e}
\usepackage[normalem]{ulem}
\usepackage{makecell}

\usepackage{physics}
\usepackage{algorithm}
\usepackage{algpseudocode}

\newcommand{\de}{\partial}
\newcommand{\dx}{\delta x}
\newcommand{\dy}{\delta y}
\newcommand{\dt}{\delta t}
\newcommand{\ds}{\delta s}

\newcommand{\cix}{c_{i,x}}
\newcommand{\ciy}{c_{i,y}}
\newcommand{\czx}{c_{0,x}}
\newcommand{\czy}{c_{0,y}}
\newcommand{\cox}{c_{1,x}}
\newcommand{\coy}{c_{1,y}}
\newcommand{\fieq}{f_i^{eq}}
\newcommand{\fione}{f_i^{(1)}}
\newcommand{\fzeq}{f_0^{eq}}
\newcommand{\foeq}{f_1^{eq}}

\begin{document}

\title{Two-dimensional quantum-lattice-gas algorithm for anisotropic Burgers-like equations}% Force line breaks with \\

\author{Niccol\'o Fonio}
\email{niccol.fonio11@gmail.com}
\affiliation{Aix-Marseille Univ, CNRS, Centrale Méditerranée, M2P2, Marseille, France }
 \affiliation{
 Aix-Marseille Univ, CNRS, LIS, Marseille, France 
}

\author{Pierre Sagaut}%
\affiliation{Aix-Marseille Univ, CNRS, Centrale Méditerranée, M2P2, Marseille, France }

\author{Giuseppe Di Molfetta}
\affiliation{
 Aix Marseille Univ, CNRS, LIS, Marseille, France 
}
\affiliation{Institute Universitaire de France, Paris, France}

\date{\today}% It is always \today, today,

\begin{abstract}
Building on hybrid quantum lattice gas algorithm, we revisit the possibilities of this quantum lattice model. By deriving a correction to the predicted viscosity, we provide analytical and numerical results that refine original formulation. We introduce a minimal 2D generalization of the algorithm, which allows to simulate anisotropic Burgers-like equations while retaining only two lattice velocities. This approach opens a promising route toward embedding momentum conservation and advancing toward Navier–Stokes dynamics in 2D, going beyond Frisch, Hasslacher and Pomeau (FHP) and lattice Boltzmann method (LBM) with a quantum native model. We highlight how the presented algorithm results more efficient in full state evolution than other quantum nonlinear solvers, nonetheless its advantages respect to classical lattice gas methods. Being this model between classical and quantum computation, it gives a unique perspective on simulating nonlinearities with quantum computers, confirming quantum lattice gas models as a crucial playground for quantum simulations of nonlinearities.
\end{abstract}
%\keywords{quantum lattice gas, quantum algorithms, nonlinearty, burger's equation}
\maketitle

\section{Introduction}
The application of quantum computing (QC) to computational fluid dynamics (CFD) seeks to uncover potential advantages over classical methods. Several paths have been explored to this end, branching in two inherently different approaches. The first strategy for nonlinear PDEs uses a linearization technique, then a block encoding \cite{low2019hamiltonian, gilyen2019quantum} or Schr\"odingerization technique \cite{jin2022quantum, hu2024quantum}, and in the end Hamiltonian simulation \cite{berry2014exponential,berry2017quantum}, charachterizing quantum nonlinear solvers (QNS). The linearization techniques mostly considered are Carleman \cite{liu2021efficient, li2025potential, itani2024quantum, sanavio2024lattice}, Koopman-von Neumann (KvN) \cite{joseph2020koopman, gan2025provably, jennings2026quantum}, and homotopy method \cite{bharadwaj2025quantum, bharadwaj2025compact}. While Carleman linearization showed some limitations for the simulable nonlinearities \cite{jennings2025end, gonzalez2025quantum, lewis2024limitations}, KvN and homotopy analysis are very promising. 
%The first regards using different linearization techniques, block encoding or schrodingerization and, in the end, Hamiltonian simulation. The algorithms developed in this sense cast the nonlinear equations in a form which is amenable to be solved with a quantum advantage. 
In particular, many of these algorithms use $O(\log(N))$ qubits to simulate discretized PDEs at $N$ points, with a query complexity usually $O(T,\log(N),h(\epsilon))$, where $h$ is the cost related to the accuracy of the algorithm $\epsilon$, possibly polynomial or polylogarithmic in $1/\epsilon$. 
%The accuracy cost for nonlinearities include the eventual limitation of the linearization method, making it depend possibly on other parameters. 
With these resources, which exclude state preparation, they achieve the evolution to a final quantum state whose amplitudes are proportional to the desired solution. Limitations arise from: (I) the linearization procedures limit the possible application span to nonlinear PDEs; (II) ultimate practical quantum advantage seems further hindered at the present time by the necessity of fault-tolerant quantum computers possible; (III) advantageous applications must regard only the measurement of observables of interest, rather then the retrieval of the state, which is inevitably inefficient.

\begin{figure}[t]
\centering
\includegraphics[width=0.9\linewidth]{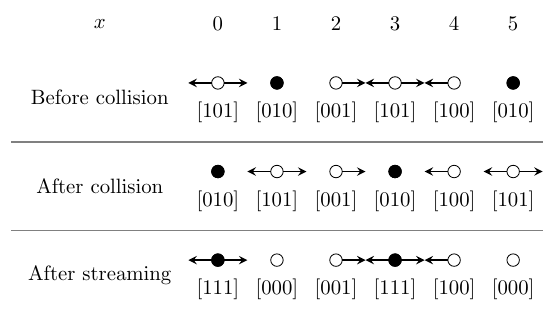}
\caption{
Evolution of a one-dimensional lattice gas cellular automaton with three discrete velocities (D1Q3). 
Each cell is represented by three bits $[b_2 b_1 b_0]$, indicating the presence of particles moving with velocities $[-1, 0, +1]$. 
The collision rule that conserves both mass and momentum (assuming the rest particle has mass 2) is $[101] \leftrightarrow [010]$, which occurs at positions $x = 0, 1, 3,$ and $5$. 
All other sites remain unaffected by the collision. 
During the streaming step, particles propagate according to their respective velocities under periodic boundary conditions.
}
\label{fig:D1Q3_lattice}
\end{figure}

The second branch for simulating nonlinear PDEs regards quantum lattice gas (QLG) models, which have an inherently different functioning.  Lattice gas models (LGM) share a common structure: a discrete lattice in which, at each site, one can define particles (binary lattice gas, BLG, or integer lattice gas, ILG) or a probability density function (lattice Boltzmann method, LBM). These undergo a local collision step followed by streaming along a discrete set of velocities. An example of the one-dimensional evolution is depicted in Fig.\ref{fig:D1Q3_lattice}.
All these methods are capable of capturing nonlinear behavior. In particular, BLG \cite{wolf2004lattice} and LBM \cite{kruger2016lattice,succi2001lattice} reproduce the Navier–Stokes equations. They avoid any linearization technique by solving the Boltzmann equation and deriving from it Navier-Stokes through a Chapman-Enskog expansion.

There are various ways to translate a classical lattice gas model to a quantum computer. One approach is to encode the LBM probability distributions into the amplitudes of a quantum state, similarly to what happens for other quantum nonlinear solvers. This amplitude encoding is widely used \cite{budinski2021quantum, ljubomir2022quantum, wawrzyniak2024unitary, kumar2025quantum}, not only for LBM but also for other LGM \cite{fonio2025adaptive, zamora2025float, wang2025quantum}. These achieve to represent a lattice of $N$ sites with $O(\log(N))$ qubits, but then generally require measurement and reinitialization at each time step for nonlinearities, which forbids any quantum advantage. This issue was partly resolved with the approach of \cite{fonio2025adaptive}, but necessitating one ancilla per time step, causing a exponentially decreasing success probability. These issues also affect linearized QLBM schemes based on Carleman’s linearization\cite{succi2023quantum, sanavio2024lattice, zecchi2025improved}, whose possibility of simulating turbulence has been analyzed in \cite{lewis2024limitations}. Multi-time-step implementations have been demonstrated a quantum advantage for advection–diffusion and other linearized models\cite{xiao2025quantum, xu2025improved, wawrzyniak2024unitary, bediche2025fully}, without achieving nonlinearities.

All these approaches consider a (probability density) function in a discretized space, and represent it with the amplitudes of a superposed quantum state, with different pros and cons. The quantum lattice gas (QLG) model we discuss, proposed by Yepez in \cite{yepez2006open}, represents a similar but conceptually different alternative. Rather than encoding a classical lattice model, it directly defines a programmable quantum lattice gas, which evolves through subsequent unitary and non-unitary operations, making it a hybrid quantum-classical method. 

For all applications where the full state is needed, QNS perform inevitably worse than classical algorithms. In this case we can ask: is there still any advantage in using quantum or quantum-inspired or hybrid models for full state evolution? The model analyzed in this article proves a positive answer. In particular, our model, compared to classical analogues (LBM), retains unconditional stability for arbitrarily small viscosities and tunable hydrodynamics parameters. The coexistance of these two elements is hard to find in any other lattice gas method, thus proving the current scheme as a promising development for quantum and quantum-inspired algorithms for fluid dynamics. Moreover, its possibilities of parallelization make it easily simulable.

While being an advanced hybrid numerical scheme for full state evolution,  this model open different perspectives. Its classical simulability, explored in \cite{love2006type}, can still be reconsidered in light of later analyses \cite{love2019quantum}, particularly when including a fully quantum streaming operator, which has not been treated yet for the model in this article. This issue is of special relevance for quantum simulations of fluid systems \cite{yepez2009vortex, yepez2009superfluid}.
Moreover, quantum walks-inspired models could show the existance of more efficient schemes for 2D simulations \cite{arrighi2018dirac,di2024quantum}, thus enabling further advantage over other classical lattice schemes.

Revisiting and extending this model, especially to two spatial dimensions, is thus of both foundational and practical interest. On one hand, it deepens our understanding of how hydrodynamic behavior emerges from a unitary, quantum-compatible lattice dynamics; on the other, it opens the possibility of constructing reduced, resource-efficient models capable of capturing 2D nonlinear dynamics. In this sense, the quantum lattice model we focus on serves not only as a testing ground for exploring the interface between quantum computation and fluid dynamics, but also as a promising pathway toward alternative mesoscopic formulations with tunable hydrodynamic parameters and lower computational costs than classical lattice gas automata, such as the Frisch–Hasslacher–Pomeau (FHP) model.

In summary, the QLG we analyze stands out as a quantum-native numerical method for simulating nonlinear dynamics. It is a conceptually different alternative to other quantum nonlinear solvers. Its possible fully quantum formulation remains unexplored, while its degree of classical simulability is still open to discussion. In this sense, the model could either enable simulations beyond classical computational capabilities, or offer classical simulations with features unattainable in existing lattice schemes, such as tunable viscosity in FHP-like models, or lower viscosities and unconditional stability compared to LBM. These considerations motivate the present work, which has two principal outcomes. In Sec.II, we derive a correction to the viscosity of \cite{yepez2006open}, providing a detailed analytical procedure elaborated in the Appendix. In Sec.III, we obtain the general partial differential equation simulated by a minimal two-dimensional extension of the model, leading to anisotropic Burgers-like dynamics. Finally, Sec.IV presents the corresponding numerical results. These can be reproduced with the related publicly available code \cite{foniocode}.

\section{Q-D1Q2}
A Q-D$n$Q$v$ model is a quantum lattice gas characterized by a lattice in $n$ dimensions where at each site there are $v$ qubits. We assume, for the sake of the algorithmic procedure we will follow, that the qubits are initially disentangled. The $k$-th qubit at site $x$ and time $t$ is in the state $\ket{q_k(x,t)}$. The cell is represented by
\begin{equation}\label{def: cell}
    \begin{split}
        \ket{\psi(x,t)} & = \bigotimes_{k=0}^{v-1} \ket{q_k(x,t)} \\
        & =\ket{q_0(x,t)}\otimes\dots\otimes\ket{q_{v-1}(x,t)},
    \end{split}
\end{equation}
and the state of the lattice at time $t$ is
\begin{equation} \label{def: lattice}
    \ket{\Psi(t)} = \bigotimes_x \ket{\psi(x,t)}.
\end{equation}
We first introduce the 1D model of \cite{yepez2006open}, since it will be the starting point for our considerations and results. We write the most important parts of the derivation, which is shown in its integrity in the Appendix.

\begin{figure*}[t]
    \centering
    \includegraphics[scale=0.65]{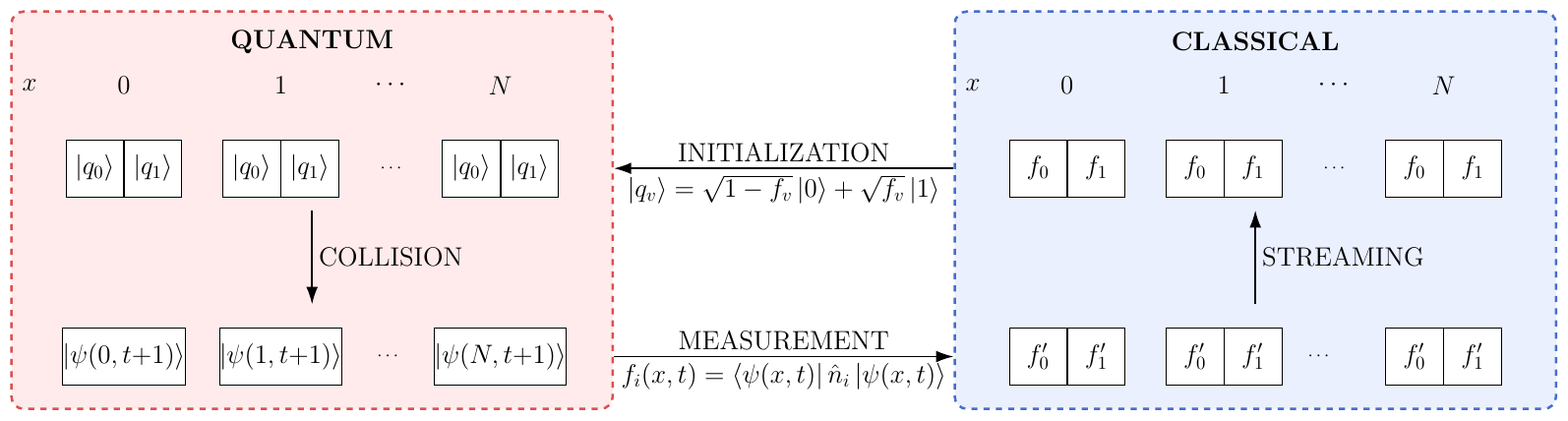}
    \caption{Algorithmic scheme for Q-D1Q2 model}
    \label{fig: algorithm}
\end{figure*}

The evolution of the system is composed of a collision, the measurement of observables of interest, and a reinitialization that carries out the classical streaming, as represented in Fig.\ref{fig: algorithm}.
The algorithmic procedure can also be written as in Alg.\ref{alg:qlgca}. The qubits are initialized as follows
\begin{equation} \label{def: qubit}
    \ket{q_v(x,t)} = \sqrt{1-f_v(x,t)} \ket{0} + \sqrt{f_v(x,t)}\ket{1},
\end{equation}
where $f_i(x,t)$ are the classical probability distribution functions of the i-th velocity. Thus, the state of each cell is 
\begin{equation} \label{eq:node state}
    \begin{split}
    \ket{\psi(x,t)} = & \sqrt{(1-f_0(x,t))(1-f_1(x,t))} \ket{00} \\ 
    & + \sqrt{(1-f_0(x,t))f_1(x,t)} \ket{01} \\ 
    & + \sqrt{f_0(x,t)(1-f_1(x,t))} \ket{10} \\
    & + \sqrt{f_0(x,t)f_1(x,t)} \ket{11}
    \end{split}.
\end{equation}
The populations $f_i$ are modified by the collision, thus they need to be measured to be updated for streaming. They result from the measurement of number operators 
\begin{equation} \label{eq:number ops}
    \hat{n}_0=\begin{pmatrix}
        0 & 0 & 0 & 0 \\
        0 & 0 & 0 & 0 \\
        0 & 0 & 1 & 0 \\
        0 & 0 & 0 & 1
    \end{pmatrix}
    \hat{n}_1=\begin{pmatrix}
        0 & 0 & 0 & 0 \\
        0 & 1 & 0 & 0 \\
        0 & 0 & 0 & 0 \\
        0 & 0 & 0 & 1
    \end{pmatrix}.
\end{equation}
having
\begin{equation} \label{eq:def fi}
    f_i(x,t) = \bra{\psi(x,t)}\hat{n}_i\ket{\psi(x,t)}.
\end{equation}
It is straightforward to verify the consistency of Eq. \ref{eq:def fi} with Eq. \ref{eq:node state}.
\begin{algorithm}[H] 
\caption{QLGCA for Burger's equation}\label{alg:qlgca}
\begin{algorithmic}
\State $\text{c\_lattice} \gets \text{initialize lattice}$ \Comment{$f_i(x,t)$}
\For {t}
\For { $x\in$ c\_lattice}
\State $\ket{\psi(x,t)} \gets \text{initialize}(f_0(x,t),f_1(x,t))$ \Comment{from Eq. \ref{eq:node state}} 
%\Comment{$\ket{\Psi(t)}=\bigotimes_x\ket{\psi(x,t)}$} 
\State $\ket{\psi(x,t)} \gets \hat{U}\ket{\psi(x,t)}$ \Comment{Collision}
\State $f_i'(x,t) \gets \bra{\psi(x,t)}\hat{U}^{\dagger}\hat{n}_i\hat{U}\ket{\psi(x,t)}$ \Comment{Measurement}
\EndFor
\For { $x\in$ c\_lattice}
\State $f_i(x,t) \gets f_i'(x-c_i,t)$ \Comment{Streaming}
\EndFor
\EndFor
\end{algorithmic}
\end{algorithm}

After the initialization, each cell undergoes a collisional operation $\hat{C}$ that must preserve the mass density of the cell, defined as $\rho(x,t)=f_0(x,t)+f_1(x,t)$. The most general unitary that ensures this conservation can be parametrized as follows
\begin{equation} \label{eq:collision op}
    \hat{C} = 
    \begin{pmatrix}
        1 & 0 & 0 & 0 \\
        0 & e^{i\xi}\text{cos}\theta & e^{i\zeta}\text{sin}\theta & 0 \\
        0 & -e^{-i\zeta}\text{sin}\theta & e^{-i\xi}\text{cos}\theta & 0 \\
        0 & 0 & 0 & 1
    \end{pmatrix},
\end{equation}
with $\theta,\zeta,\xi$ being real angles, commonly said Euler's angles. After the collision, $\hat{n}_i$ are measured and give the post-collisional populations $f_i'(x,t)$. Considering also the streaming, which is classical, we know that $f_i(x-c_i \delta x,t+\delta t)=f_i'(x,t)$, being $\delta x$ the lattice spacing, $\delta t$ the time step, $c_i$ the corresponding velocity. We can then write the evolution equation
\begin{equation} \label{eq: finite_diff_1d}
    f_i(x-c_i \delta x,t + \delta t) = f_i(x,t) + \Omega_i(x,t),
\end{equation}
where $\Omega_i$ is the collision term. $\Omega_i$ in general depends on $f_i(x,t)$. The equilibrium condition is the equation $\Omega_i=0$, which allows us to calculate $f_i^{eq}$. In quantum terms, we write
\begin{equation}
    \Omega[\psi(x,t)] = \bra{\psi(x,t)}\hat{C}^{\dagger}\hat{n}_i \hat{C} - \hat{n}_i \ket{\psi(x,t)}.
\end{equation}
For this model, given by the mass conservation imposed by the collision, we have $-\Omega_0=\Omega_1=\Omega$, where
\begin{equation}
    \begin{split}
        \Omega = & (f_0-f_1)\sin^2\theta + \\
        & +\sin(2\theta)\cos(\zeta-\xi)\sqrt{f_0(1-f_0)f_1(1-f_1)}.
    \end{split}
\end{equation}
Solving the equilibrium condition brings us to
\small
\begin{equation} \label{eq:equilibrium_1D}
    \begin{split}
    f_0^{eq}&=\frac{\rho}{2} + \frac{1}{2\alpha}\bigg[ \sqrt{\alpha^2+1}-\sqrt{(\alpha^2+1) - 2\alpha^2\rho + \alpha^2\rho^2} \bigg] \\
    f_1^{eq}&=\frac{\rho}{2} - \frac{1}{2\alpha}\bigg[ \sqrt{\alpha^2+1} -\sqrt{(\alpha^2+1) - 2\alpha^2\rho + \alpha^2\rho^2}\bigg],
    \end{split}
\end{equation}
\normalsize
where $\alpha=\cot \theta \cos(\zeta-\xi)$. It is very convenient to obtain the closed-form solution of this QLG model. We will use this feature for the generalization of the model to 2D. We stress out, in particular, that the expressions in Eqs. \ref{eq:equilibrium_1D} depend on the collision model and upon the fact that the populations are not mixed, thus $f_i$ streams to $f_i$. In the case of mixed streaming, e.g. $f_0$ goes to $f_1$, the equilibrium would be affected. 

To obtain the PDE we are simulating with this model, we start from Eq. \ref{eq: finite_diff_1d}. This will also be the starting point for the 2D derivation. We first do the Taylor expansion of the finite difference up to the second order in $\delta x$, and first order in $\delta t$. This is due to the fact that we will find ourselves in the diffusive regime where $\delta x \approx \epsilon$ and $\delta t \approx \epsilon^2$ where $\epsilon$ is the Knudsen number. After the Taylor expansion, we do the Chapman-Enskog expansion. At the first-order in $\epsilon$, we can solve the corresponding equation to find $f_i^{(1)}$ as an expression of $f_i^{eq}$. In 1D we obtain
\begin{equation} \label{eq: sol_fione}
    \epsilon f_i^{(1)} = -\frac{1}{J_1-J_0}c_i \delta x \partial_x f_i^{eq}
\end{equation}
with $J_i=\pdv{\Omega}{f_i}\big|_{eq}$. In the appendix, we show explicitly how to calculate $J_1-J_0$, which results in  
\begin{equation}\label{eq:deltaJ}
    J_1-J_0 = -2\sin^2(\theta) \qty[1 - \alpha^2(1-\rho) - \alpha^2\frac{\rho^2-u^2}{2}].
\end{equation}
Including also the second order, we obtain a PDE for each $f_i$ as follows 
\begin{equation} \label{eq:pde_fieq}
    \delta t\partial_t f_i^{eq} - c_i \delta x \de_x (f_i^{eq}+ \epsilon f_i^{(1)}) + \frac{\delta x^2}{2} \partial_{xx}f_i^{eq} + O(\epsilon^3) = \Omega_i.
\end{equation}
This allows us to find a clear PDE for the mass density summing Eq. \ref{eq:pde_fieq} for $(i=0)$ and $(i=1)$. The collision term in the RHS vanishes due to the conservation of mass. We obtain
\begin{equation}
    \begin{split}
    \delta t\partial_t \rho - & \delta x \de_x [f_1^{eq}-f_0^{eq} +  \epsilon(f_1^{(1)}-f_0^{(1)})] + \\ 
    + &\frac{\delta x^2}{2} \partial_{xx}\rho + O(\epsilon^3) = 0.
    \end{split}
\end{equation}
Defining $u = f_1^{eq}-f_0^{eq}$ and remembering Eq. \ref{eq: sol_fione}, we have
\begin{equation}
    \begin{split}
        \delta t\partial_t \rho &-  \delta x \de_x u +   \delta x^2 (\de_x \rho) \de_x \qty(\frac{1}{J_1-J_0})  + \\
        &+ \frac{\delta x^2}{2} \qty(1+\frac{2}{J_1-J_0}) \partial_{xx}\rho + O(\epsilon^3) = 0.
    \end{split}
\end{equation}
We can now rewrite the PDE only in terms of $\rho$, having
\small
\begin{equation} \label{eq:pde_1d_fulldj}
    \begin{split}
    &\delta t\partial_t \rho - \delta x \frac{\alpha(\rho-1)\de_x\rho}{\sqrt{1+\alpha^2(\rho-1)^2}} +\\
    &+\delta x^2 (\de_x \rho) \frac{1}{2 \sin^2(\theta)} \frac{\alpha^2 [-\de_x\rho + \rho\de_x\rho-u\de_xu]}{(1+\alpha^2(1-\rho) + \alpha^2\frac{\rho^2-u^2}{2})^2}  + \\ 
    &+\frac{\delta x^2}{2} \qty(1-\frac{1}{2 \sin^2(\theta)} \frac{1}{1+\alpha^2(1-\rho) + \alpha^2\frac{\rho^2-u^2}{2}}) \partial_{xx}\rho \\ 
    & + O(\epsilon^3) = 0.
    \end{split}
\end{equation}
\normalsize
The next step consists of considering $\rho-1 = O(\epsilon)$. It is motivated at this point by the presence of terms $\propto1/(\rho-1)^a$ for some $a$, which can clearly affect the considered order in $\epsilon$. Other incompressible regimes with $\rho \approx \rho_0$ are of lesser interest. The correspondance between this approximation and the low-Mach regime will be clear after the change of variable.

The explicit calculation for this incompressible regime can be found in the appendix, while here we just remind the important fact that we use $O((\rho-1))=O(\epsilon)$. This allows us to obtain the following equation
\begin{equation} \label{eq:final_pde_1d}
    \begin{split}
        \dt &\partial_t \rho  + \dx \alpha(1-\rho)\de_x \rho +\\
        &+ \frac{\delta x^2}{2} \qty(1-\frac{1}{\sin^2(\theta)\sqrt{\alpha^2+1}}) \partial_{xx}\rho + O(\epsilon^4) = 0.
    \end{split}
\end{equation}
In \cite{yepez2006open}, the same equation is obtained but with a different viscosity, in particular
\begin{equation} \label{eq:final_yepez_1D}
     \begin{split}
         \dt & \partial_t \rho + \dx \alpha(1-\rho)\de_x \rho  \\ 
         &- \frac{\delta x^2}{2}\cot^2(\theta) \partial_{xx}\rho + O(\epsilon^3,\alpha^2\epsilon) = 0
     \end{split}
\end{equation}
This difference is due to the approximation for $J_1-J_0$, requiring the correctness of the above equation only for $O(\epsilon^3,\epsilon\alpha^2)$.

Since every term is at the same order, we can consider the formal limit of Eq. \ref{eq:final_pde_1d} and write the equation followed by any one-dimensional mass-preserving quantum lattice gas as defined
\begin{align}
    \de_t \rho +c_s(1-\rho)\de_x\rho=\nu\de_{xx}\rho \label{eq:final_pde_1d_bis}\\
    c_s = c\alpha=\frac{\dx}{\dt}\cot(\theta)\cos(\zeta-\xi) \label{eq:cs_1d}\\
    \nu = - \frac{\dx^2}{\dt}\frac{1}{2} \qty(1-\frac{1}{\sin^2(\theta)\sqrt{\alpha^2+1}}). \label{eq:nu_1d}
\end{align}
We can connect this equation to Burger's equation with a change of variable. Using 
\begin{equation} \label{eq:change_of_variables}
    w = c_s(1-\rho),
\end{equation}
it is straightforward to see that we obtain
\begin{equation}
    \de_t w+w\de_x w=\nu\de_{xx}w,
\end{equation}
which is exactly the Burger's equation. For the change of variable in Eq. \ref{eq:change_of_variables}, taking the limit for $\rho-1\approx 0$ corresponds to taking the limit $w/c_s\approx 0$, and this is well known as the low-Mach regime. Considering another incompressible regime $\rho\approx\rho_0$ would not simulate Burger's equation in low-Mach. We point out that $w$ corresponds to a macroscopic velocity field, but it is not identical to the microscopic velocity field.

This result was possible thanks to a series of factors. First of all the collision allowed for calculating the exact microscopic expression of the equilibrium populations. This is not generally guaranteed, and usually in LGCA it is necessary to apply an approximation to get their expression \cite{wolf2004lattice} as function of mass and momentum. In the second place, we notice that the expressions we got for $J_1-J_0$, which was fundamental, depends on the collision we apply. This means that the 2D generalization that we propose, as any model having the same collision and streaming structure can then inherit the same equilibrium distributions and the same $J_1-J_0$.

\section{Q-D2Q2}
In this section, we introduce a generalization of the presented QLG model to 2 dimensions. In particular, 2D QLGA have already been studied with FHP model in \cite{micci2015measurement}
, where fluid dynamic simulations were carried out. We are interested, instead, in studying the behavior of QLGA on a square lattice where, at each site, there are 2 qubits, in order to obtain the 2D Burger's equation. Using only two velocities will inevitably bring us to anisotropic equations. However, this will allow for a complete analytical result without considering further approximations, extending naturally the Burger-like behavior to 2D. 

For simplicity we assume that the qubits are arranged in a square grid. We will see that our analytical result can be valid on any Bravais lattice in 2D. Each pair of qubits is initialized as in the 1D model and undergo the same collision. We allow the populations to stream also in the $y-$direction. We keep the same number of qubits per site and collision because, as we said, a closed-form solution for the equilibrium distributions is convenient. 

We start from the following expression, given by the algorithmic procedure, considering for simplicity $\dx=\dy = \ds$,

\begin{equation} \label{eq: finite_diff_2d}
    f_i(x - \cix \ds, y - \ciy \ds, t+ \dt) - f_i(x,y,t) = \Omega_i.
\end{equation}

We do the Taylor expansion, remembering the diffusive scaling $\ds \approx\epsilon$ and $\dt \approx\epsilon^2$, and the Chapman-Enskog expansion, obtaining

\begin{widetext}
\begin{equation}\label{eq: taylor_2d_1}
    \begin{split}
        f_i(x- \cix \ds,  y - \ciy &\ds,  t+ \dt) - f_i(x,y,t) = 
         \dt \de_t \fieq - \epsilon \ds (\cix \de_x \fione + \ciy \de_y\fione) + \\
        -& \ds (\cix \de_x \fieq + \ciy \partial_y \fieq) + \frac{\ds^2}{2}(\cix^2\de_{xx}\fieq + \ciy^2\de_{yy}\fieq + \\ 
        +& 2\cix \ciy \de_{xy}\fieq).
    \end{split}
\end{equation}
\end{widetext}
Eq. \ref{eq: taylor_2d_1} is the LHS of Eq. \ref{eq: finite_diff_2d}. The RHS, since it is not affected by the streaming because it depends on $f_i(x,t)$, is the same as in 1D. We solve the equation at the first order, which now depends on $c_i$'s, obtaining 

\begin{widetext}
    \begin{equation} \label{eq: cione}
    \begin{split}
        \cix \de_x \fione + \ciy \de_y \fione = & -\frac{1}{J_1-J_0}(\cix^2 \de_{xx} \fieq + 2\cix\ciy\de_{xy} \fieq + \ciy^2 \de_{yy} \fieq) + \\
        & - (\cix \dx \fieq + \ciy \dy \fieq) \ds (\cix \de_x + \ciy \de_y)\frac{1}{J_1-J_0}.
    \end{split}
\end{equation}
\end{widetext}

Now we substitute Eq. \ref{eq: cione} in Eq. \ref{eq: taylor_2d_1} and we sum the two equations for $(i=0)$ and $(i=1)$. We sum the two equations because we are interested in studying the behavior of the conserved quantity of interest, the mass density in this case. For this quantity, we notice as before that the RHS is 0, because we chose the collision as in 1D to have $-\Omega_0 = \Omega_1$. 

We can then rewrite this expression using $\foeq = \frac{\rho + u}{2}$ and $\fzeq = \frac{\rho - u}{2}$, with $u=f_1-f_0$. We notice that $u$ in this case is just a convenient variable, it is not connected necessarily to the momentum in any direction. 

We then apply the incompressible approximation as before. Since we haven't modified the equilibrium distributions, the definition of $u$ does not change, as well as the definition of $J_1-J_0$. The incompressible approximation always consists of $\rho\approx1$, and the Eq. \ref{eq:final_pde_2d} is valid in that limit

\small
\begin{equation} \label{eq:final_pde_2d}
    \begin{split}
        \de_t \rho + & \frac{c}{2} [(\czx+\cox)\de_x \rho + (\czy+\coy)\de_y \rho]
        + \\
        + & \frac{c_s}{2} (1-\rho) [ (-\czx+\cox)\de_x \rho + (-\czy+\coy)\de_y \rho ]+ \\
        - & \frac{\nu}{2} [ (\czx^2+\cox^2)\de_{xx}\rho + (\czy^2+\coy^2)\de_{yy}\rho \\&+ 2(\czx\czy+\cox\coy) \de_{xy}\rho ] + O(\epsilon^4) = 0.
    \end{split}
\end{equation}
\normalsize
If we define the constant vectors and anisotropic diffusion tensor as follows
\begin{align}
    \mathbf{a} &= \tfrac{c}{2}(\czx+\cox,\; \czy+\coy),\\
    \mathbf{b} &= \tfrac{c_s}{2}(\czx-\cox,\; \czy-\coy),\\
    D &= \tfrac{\nu}{2}
    \begin{pmatrix}
        \czx^2+\cox^2 & \czx\czy+\cox\coy\\[4pt]
        \czx\czy+\cox\coy & \czy^2+\coy^2
    \end{pmatrix}.
\end{align}
Then Eq. \ref{eq:final_pde_2d} can be written compactly as
\begin{equation} 
    \partial_t \rho
    + \mathbf{a}\!\cdot\nabla\rho
    + \mathbf{b}\!\cdot\!\big[(1-\rho)\nabla\rho\big]
    - \nabla\!\cdot(D\nabla\rho)
    = 0.
    \label{eq:pde_vector}
\end{equation}
or
\begin{equation} 
    \partial_t \rho
    + \mathbf{d}\!\cdot\nabla\rho
    - \nabla\!\cdot(D\nabla\rho)
    = 0, 
    \label{eq:pde_vector-bis}
\end{equation}
with
\begin{equation} 
 \mathbf{d} = \mathbf{a} + (1-\rho) \mathbf{b}.
    \label{eq:pde_vector-ter}
\end{equation}
This is the final expression we obtain, thus the generalization to 2D for the Q-D1Q2 model, in the limit $\rho-1\approx0$. The change of variables, and consequent connection to low-Mach regime, can be easily found as before. In the first place, we can see that this includes the 1D case. In fact for $\Vec{c}_0=(-1,0)$ and $\Vec{c}_1=(1,0)$, we obtain Eq. \ref{eq:final_pde_2d}. Secondly, we can notice an additional advective term. This appears only when the velocities are not symmetric, as we would expect. We also notice that there is a cross-term proportional to $\de_{xy}\rho$ that can disappear for specific choices of $\vec{c}_i$. The non-linear term is preserved. This disappears in the trivial case $\vec{c}_0=\vec{c}_1$.

\begin{figure*}
    \centering
    \includegraphics[width=0.75\linewidth]{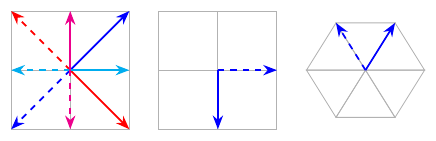}
    \caption{Different possible velocity sets. Continuous (dashed) arrows correspond to $f_1$ ($f_0$) streaming direction. On the left we have $c_{0,i}=-c_{1,i}=c_i$. At the center we have $c_{0,x}=1$,$c_{0,y}=0$,$c_{1,x}=0$,$c_{1,y}=-1$. On the right we have a triangular lattice and $\vec{c}_0=(-1/2,\sqrt{3}/2)$ and $\vec{c}_4=(1/2,\sqrt{3}/2)$}
    \label{fig:velocity schemes}
\end{figure*}

This expression is valid for an arbitrary discrete velocity set that can identify a Bravais lattice. Thus, we can use any 2D Bravais lattice. We consider, as an example, the different choices represented in Fig.\ref{fig:velocity schemes}.
We first consider a square lattice. We can have $c_{0,i}=-c_{1,i}=c_i$, obtaining
\begin{equation}\label{eq:rho_cs_nu}
    \begin{split}
        \de_t \rho + &c_s (1-\rho) [ 2c_x\de_x \rho + 2c_y\de_y \rho ] \\
        - & \nu [ 2c_x^2\de_{xx}\rho + 2c_y^2\de_{yy}\rho + 4c_xc_y \de_{xy}\rho ] = 0.
    \end{split}
\end{equation}
However, this is just a 1-dimensional dynamics. There is no real mixing of information in different directions. Thus, the appearance of terms in x and y is just a matter of frame of reference choice.

We can get rid of the cross-derivatives term imposing $\czx\czy+\cox\coy=0$. For example we could define $c_{0,x}=1$,$c_{0,y}=0$,$c_{1,x}=0$,$c_{1,y}=-1$, thus having the population $f_0$ moving horizontally and the population $f_1$ moving vertically. We then obtain
\begin{equation}  \label{eq:third velocity}
    \de_t \rho +  c  [ \de_x \rho -\de_y \rho ] + c_s (\rho-1)\nabla \rho   = \nu \nabla^2\rho.
\end{equation}
This is the Burger's equation with an additional advective term. Thus, we can remove the anisotropy, but we add an advective term.

On a triangular lattice if we chose, for example, $\vec{c}_0=(-1/2,\sqrt{3}/2)$ and $\vec{c}_1=(1/2,\sqrt{3}/2)$, then we would obtain
\begin{equation}\label{eq:rho_simple}
    \begin{split}
        \de_t \rho + \frac{c}{2} \sqrt{3}\de_y \rho &+ c_s' (\rho-1) 2\de_x \rho \\
        &+ \nu' [ 2\de_{xx}\rho + \frac{3}{2}\de_{yy}\rho]  = 0.
    \end{split}
\end{equation}
Also in this case we have an additional advective term, but only in the y-direction. We got the non-linear term only in the x-direction, and the diffusive term is anisotropic, but it does not involve any cross derivatives.

All these velocity sets show that it is possible to obtain a Burger-like nonlinear PDE for this QLG in 2 dimensions. These equations can show some additional anisotropies or advective terms depending on the discrete set of lattice velocities, and also on the lattice itself. Since Eq. \ref{eq:final_pde_2d} is the most general result we can obtain with 2 velocities, the features we saw, like the anisotropies or the advective terms, may be overcome only by adding more qubits. This would completely change the model, since it would need to define an appropriate mass-conserving collision. A 3-velocity model may also be useful for introducing momentum conservation. This is left as a future perspective.

\section{Numerical results}

\subsection{Viscosities in 1D}
\begin{figure}[t]
    \centering
    \includegraphics[width=\linewidth]{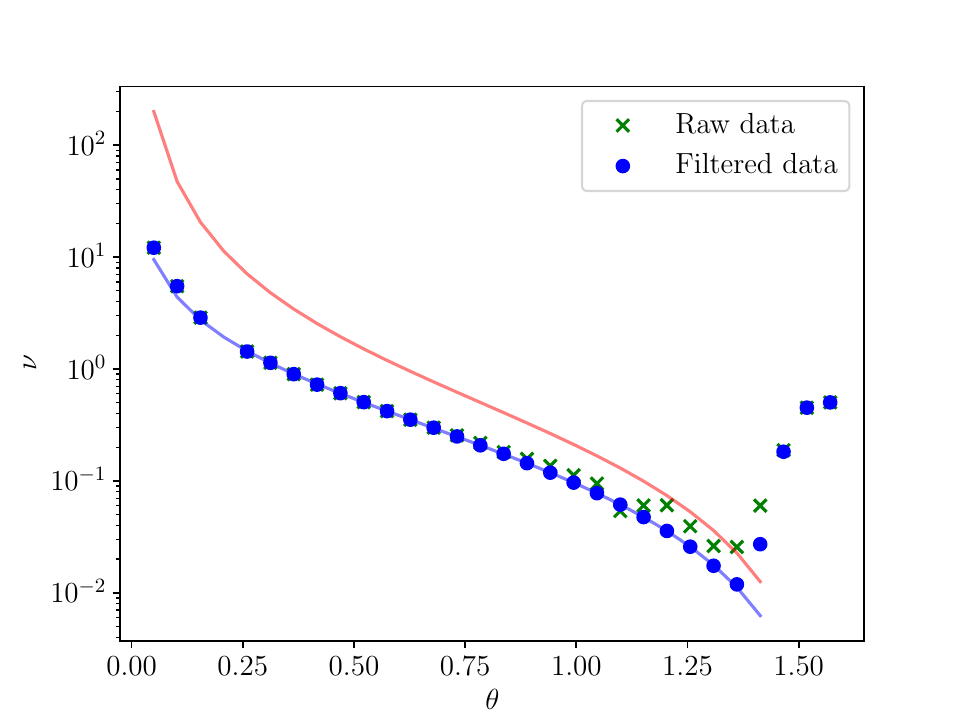}
    \includegraphics[width=\linewidth]{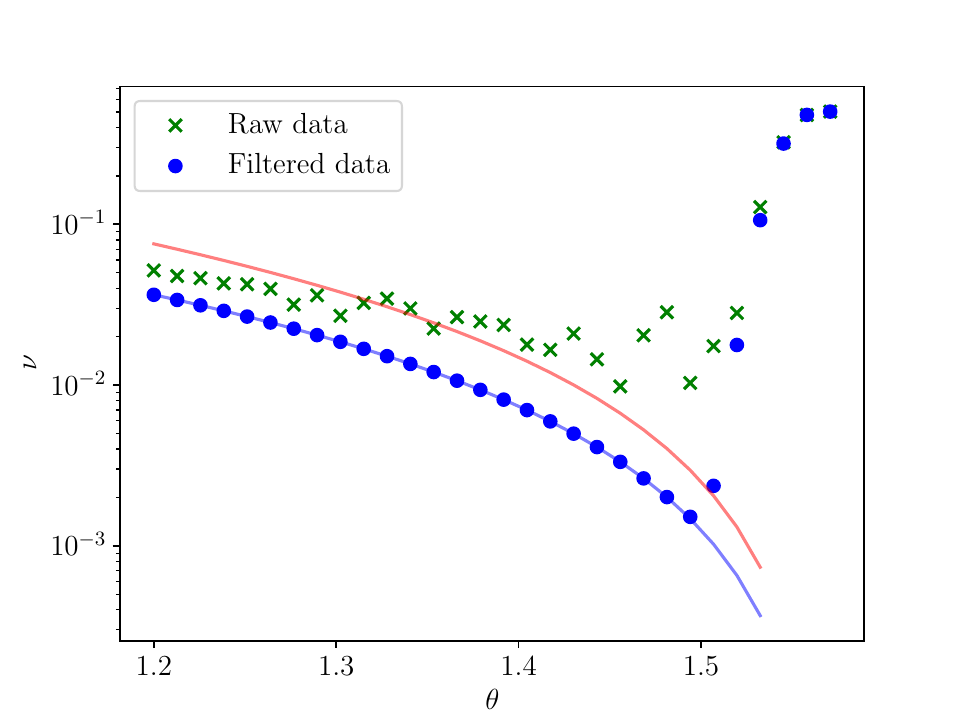}
    \caption{Simulation viscosities calculated according to Eq. \ref{eq:experimental_nu}, initializing the system with $N_x=64$ and $\rho_a=0.005$, doing the average for $T=200$ timesteps for the plot above, and $T=2000$ for the plot below. The blue line is the viscosities in Eq. \ref{eq:nu_1d}, while the red line is the viscosity of \cite{yepez2006open} $\cot^2(\theta)/2$}
    \label{fig:viscosities}
\end{figure}

To compare the viscosities, we run simulations for different $\theta$s, extract the viscosity of this system numerically,  and plot the result. The viscosity in the simulations is calculated as
\begin{widetext}
\begin{equation}\label{eq:experimental_nu}
    \nu_{exp}(t) = \sum_x\frac{1}{N_x} \frac{\rho(x,t+1)-\rho(x,t) + \alpha(\rho-1)(\rho(x+1,t)-\rho(x,t))}{\rho(x-1,t)-2\rho(x,t)+\rho(x+1,t)}.
\end{equation}
\end{widetext}
For validating our hypothesis numerically, it is necessary to reduce the numerical instability caused by the shock. For this, we chose a regime with $\rho_a=0.005$. Even with this initial condition, the presence of the shock could affect the average process. Thus we follow this procedure: (1) simulate the system, (2) look at the viscosity in space for each time step, (3) filter out the points away from 1 standard deviation with respect to the space average for each time step, (4) average  the filtered data over space, (5) average the filtered data over time. From Fig.\ref{fig:viscosities}, we see that the simulation results follow the viscosity we predicted.

The simulations where carried out for $\theta\in[0.05,\frac{\pi}{2}]$, having 30 values. In fact, we see that $\alpha \propto1/\sin(\theta)$ diverges for $\theta=0$. Filtering was more effective for increasing angles after $\theta\approx1$. 

However, even filtered data could not explain the disagreement with the tale. For this, we need to get deeper into the inviscid regime. In fact, for $\theta \approx 0$ we have a collision with a small effect on the populations, causing them to change very slowly. For $\theta=0$ populations are not affected by the collision, avoiding thermalization, thus requiring infinite time to align with our prediction. For $\theta=\pi/2$ we have populations exchange, which also cannot thermalize. However, data are more stable because there is no divergence of $\alpha$. Thus, the last points in the tale necessitate more time to get to the equilibrium. If we run the same simulation for $T=2000$ time steps and using $\theta\in[1.2,\pi/2]$, we obtain the result in Fig.\ref{fig:viscosities}.

We align more points to our prediction, reaching agreement for $\theta\approx1.5$. This shows that eventual disagreement in the inviscid regime can be due to longer relaxation times. The same effect can be seen considering the same simulations as Yepez, thus having $\rho_a=0.4$.

\subsection*{Arbitrarily small viscosities}

In single-relaxation-time lattice Boltzmann (LBM) models, the kinematic viscosity is given by
\begin{equation}
    \nu_{\text{LBM}} = c_s^2 \left(\tau - \frac{1}{2}\right),
\end{equation}

which imposes a lower bound on the attainable viscosity due to numerical stability constraints: single-relaxation-time LBM becomes unstable as $\tau \to 0.5$ \cite{wissocq2019extended,wissocq2022hydrodynamic}. Hence, the viscosity cannot be reduced arbitrarily. However, it can be controlled. As an alternative, Entropic lattice Boltzmann models provide a more stable numerical method because they satisfy an H-theorem \cite{hosseini2023entropic}. However, in these models, you cannot usually control the viscosity. 

In contrast, the presented QLG model follows an H-theorem, as proven in \cite{yepez2006open}, thus making it unconditionally stable. Additionally, the viscosity can be controlled as explained by Eq. \ref{eq:nu_1d}
\begin{equation*}
    \nu_{\text{QLG}} = -\frac{\dx^2}{\dt} \frac{1}{2}
\left( 1 - \frac{1}{\sin^2(\theta)\sqrt{\alpha^2 + 1}} \right).    
\end{equation*}
It follows that the viscosity can be made arbitrarily small. However, as discussed in the previous section, the collision strength (controlled by $\theta$) influences the relaxation dynamics. For $\theta \approx 0$ or $\theta \approx \pi/2$, populations thermalize more slowly, delaying shock formation. This causes longer simulation times for smaller or higher viscosities

\begin{figure}[b]
\centering
\includegraphics[width=0.95\linewidth]{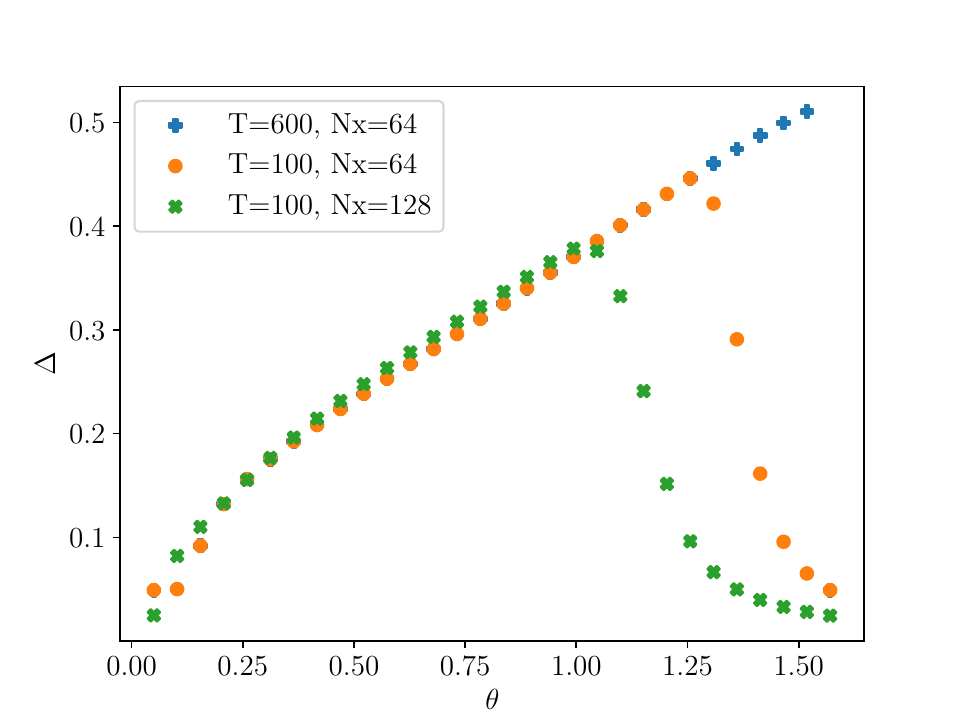}
\caption{Maximal spatial gradient of $w(x,t)/\alpha$ for different parameters.}
\label{fig:steepness}
\end{figure}

To illustrate this effect, we computed the maximal steepness of the shock front by evaluating the spatial gradient of $\tilde{w} =w(x,t)/\alpha$ [Eq. \ref{eq:change_of_variables}] and tracking its peak value over a fixed number of time steps. In particular, we calculated numerically
\begin{equation}
    \Delta = \max_{x,t}|\tilde{w}(x,t)-\tilde{w}(x+1,t)|.
\end{equation}
Since the steepness is inversely related to the effective viscosity, sharper fronts correspond to smaller viscosities. The results are shown in Fig.\ref{fig:steepness}.

Several trends emerge. Increasing $\theta$ initially enhances the steepness, corresponding to a decrease in viscosity. Near $\theta \approx \pi/2$, however, the steepness decreases because the shock has not yet fully developed—the relaxation is too slow. Comparing the curves at different times (blue square crosses and orange points) confirms that the front continues to sharpen with longer evolution times.

\begin{figure*}
    \centering
    \includegraphics[width=\linewidth]{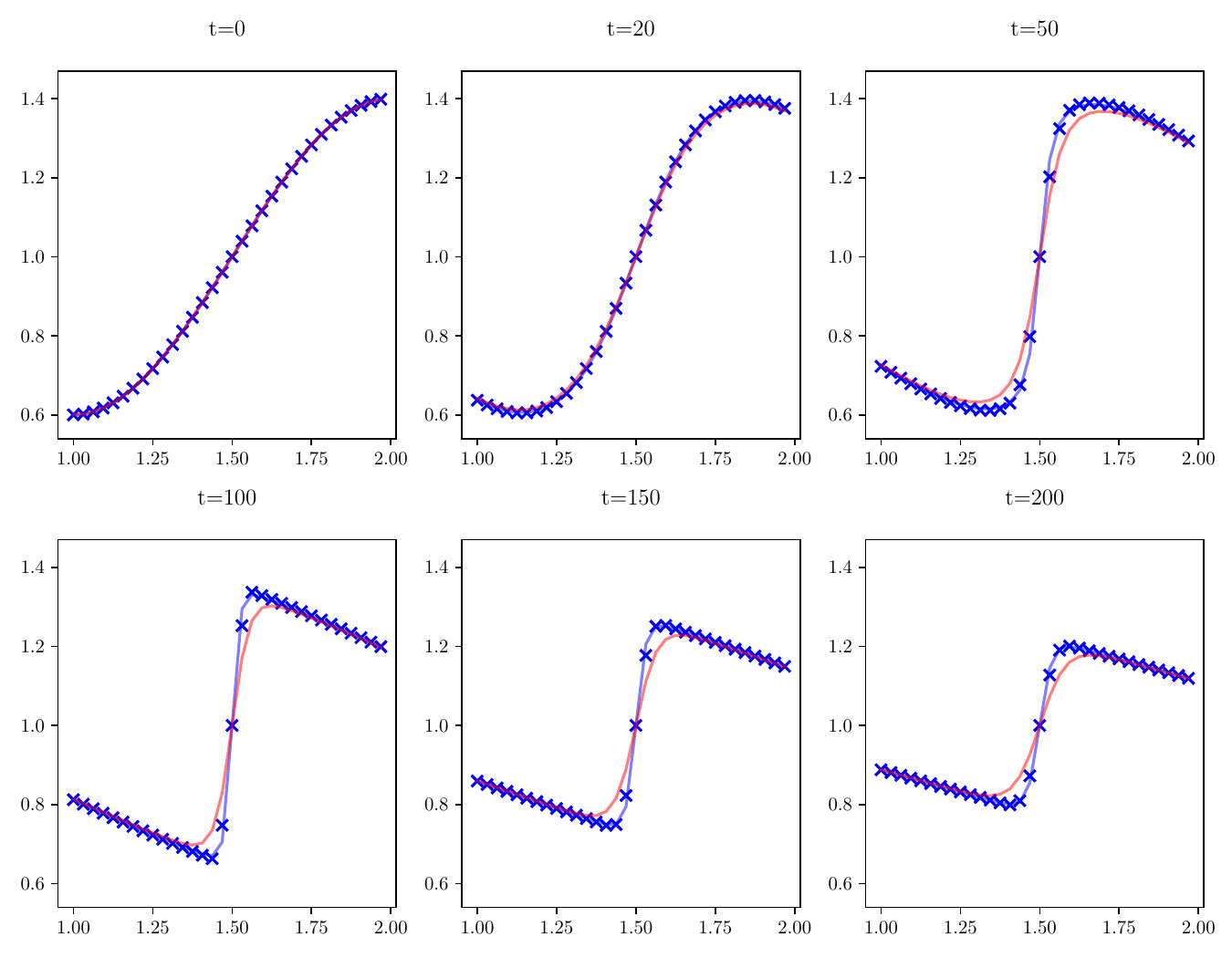}
    \caption{Simulation with $N_x=64$, $L_x=2$, $\rho_a=0.4$, $\rho_b=1$, $\theta=\pi/3$, $\zeta=\xi=0$ at different time steps. Comparison with the analytical solution with viscosity as in Eq. \ref{eq:nu_1d}(blue line), the analytical solution with viscosity as in Eq. \ref{eq:final_yepez_1D} (red line), and data from QLGCA simulation (blue crosses). We simulated the system with initial condition as in Eq. \ref{eq:initial_1D}, but for visual clarity we depicted only points for $x\in[1,2]$.}
    \label{fig:analytical comparison}
\end{figure*}

Finally, Eq. \ref{eq:nu_1d} indicates that the viscosity scales with both spatial and temporal discretization. Consistently, simulations with finer grids exhibit higher effective viscosities (green points).
Overall, these results demonstrate that the QLG model can achieve arbitrarily small viscosities without compromising numerical stability, at the cost of longer equilibration times or reduced spatial resolution.

\subsection{Analytical comparison}

We can now compare the viscosities in Eq. \ref{eq:final_pde_1d_bis} and Eq. \ref{eq:final_yepez_1D} against the analytical solution. Our equation can be mapped to a Burger's equation which is analytically solvable. We use the same method as in \cite{yepez2006open}. We start from Eq. \ref{eq:final_pde_1d_bis}
\begin{equation*}
    \de_t \rho +c_s(1-\rho)\de_x\rho=\nu\de_{xx}\rho.
\end{equation*}
We use periodic boundary condition over the domain $x\in[0,L_x)$, having the initial condition
\begin{equation} \label{eq:initial_1D}
    \rho(x,0) = \rho_b + \rho_a\cos \qty(\beta x),
\end{equation}
with $\beta = 2\pi/L_x$. We do apply the change of variable in Eq. \ref{eq:change_of_variables}, $w=c\alpha(1-\rho)$, thus having the initial condition
\begin{equation} \label{eq:initial_condition_burger}
    w(x,0)=c\alpha[1-\rho_b-\rho_a\cos(\beta x)] = w_0(x).
\end{equation}
We remember that $c=\dx/\dt$ and $\dx=L_x/N_x$ and $\dt=\dx^2$. We can solve the Burger's equation with initial condition in Eq. \ref{eq:initial_condition_burger} using the Cole-Hopf transformation
\begin{equation} \label{eq:cole-hopf}
    w(x,t) = \Bar{w} - 2\nu \de_x \ln\psi,
\end{equation}
with 
\begin{equation}
    \bar{w}=\frac{1}{L_x} \int_{0}^{L_x} w_0(s) d s = c\alpha(1-\rho_b).
\end{equation}

This transformation turns the Burger equation into a heat equation for $\psi(x,t)$. The choosen initial condition makes it profitable to write the solution with modified Bessel functions $I_i$. For general initial conditions, Fourier coefficients for the heat equation should be calculated. Thus, the solution with our initial condition is explicitly
\begin{equation}
    \psi(x,t) = I_0(A) + 2\sum_{l=1}^{\infty} I_l(A)e^{-\nu l^2 \beta^2 t} F_l(x),
\end{equation}
with
\begin{equation}
    F_l(x) = \cos(\frac{l\pi}{2}) \cos(l\beta x) + \sin\qty(\frac{l\pi}{2}) \sin(l\beta x),
\end{equation}
where $A=\frac{c\alpha\rho_a}{2\nu\beta}$. From here we can calculate $w(x,t)$ using Eq. \ref{eq:cole-hopf} and then $\rho(x,t)$ with Eq. \ref{eq:change_of_variables}. To obtain the same condition as Yepez, we run the simulation truncating the series to 80 terms. We simulate the systems with the parameters specified in Fig.\ref{fig:analytical comparison}.

\begin{figure}
    \centering
    \includegraphics[width=\linewidth]{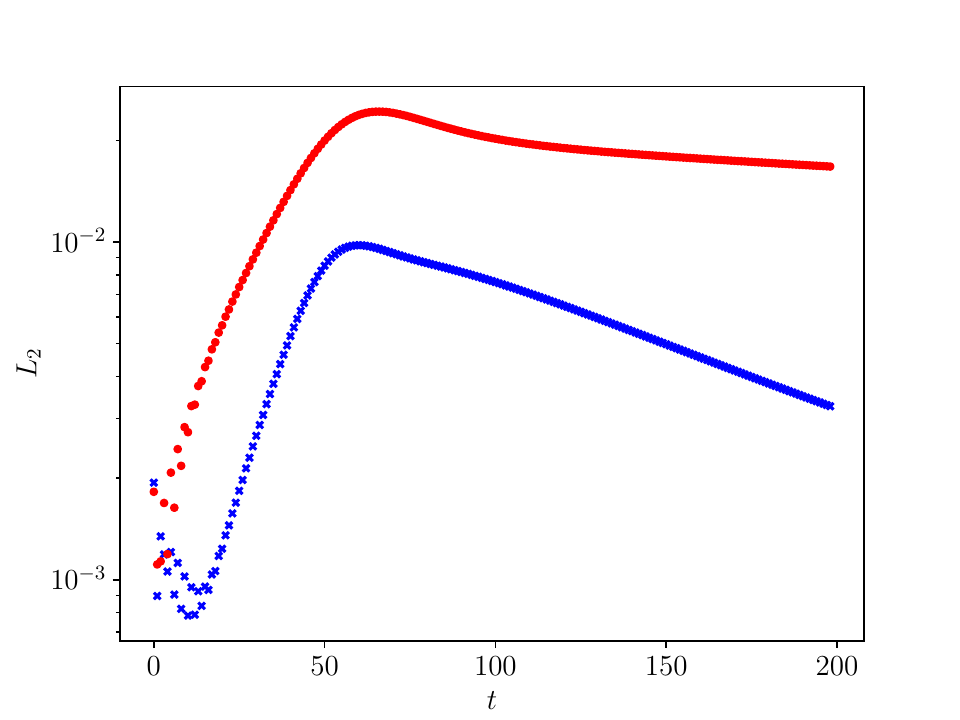}
    \caption{The red dots represent the mean square error between the analytical solution with viscosity of Eq. \ref{eq:final_yepez_1D} and the simulation data. The blue crosses represent the mean square error between the analytical solution with viscosity of Eq. \ref{eq:final_pde_1d_bis} and the simulation data.}
    \label{fig:L2_comparison}
\end{figure}

We can see that the analytical solution calculated with the corrected viscosity reproduces better the quantum lattice gas. This explains also the small gap that was already reported in \cite{yepez2006open}. This comparison can be carried out by calculating the mean squared errors, having the results in Fig.\ref{fig:L2_comparison}. Here we see the error decreasing because of the damping of the wave occurring after the formation of the shock.

\subsection{Simulations in 2D}

We initialize the system with the following initial condition
\begin{align}
    \rho(x_i,y_j,0)&=\rho_b+\rho_a \qty[\cos \qty(\frac{2\pi i}{N_x}) + \cos \qty(\frac{2\pi j}{N_y})] \label{eq:initial_cos}. %\\
    %\rho(x_i,y_j,0)&=\rho_b+\rho_a \qty[\cos \qty(\frac{2\pi i}{N_x}) \cos \qty(\frac{2\pi j}{N_y})] \label{eq:initial_cross_cos}
\end{align}
In Fig.\ref{fig:2D_simulations} we can see the time evolution of the quantum lattice gas for different time steps.

In particular, we can appreciate the shock formation in each of the 4 cases. The first two are actually equivalent to 1D dynamics, while the third and fourth velocity sets actually mix the information along two orthogonal axes. We can still appreciate the shock formation with the third velocity set, as predicted in Eq. \ref{eq:third velocity} while having an advection. In the fourth case, it is evident as well the shock formation and the advection, with an additional asymmetry brought by the asymmetry of the velocity set.

In this article, we do not give an explicit solution for the 2D equation we found. However, its correctness can be assessed with the comparison of the same equation solved by another numerical method. Thus, we consider Eq. \ref{eq:final_pde_2d} and solve it with an explicit Finite Difference-based numerical method. As we can see in Fig.\ref{fig:2D_simulations_withFDM_2} there is a good qualitative and numerical agreement between the quantum lattice gas and the Finite Difference method solution. The QLG was, as expected, more stable, and the comparison with $L_2$ norm is shown only until the FDM did not explode. This usually occurred, as expected, at the formation of the shock, which can be attributed to as the cause of the instability of the FDM adopted. The agreement, however, did not motivate a further comparison with other classical numerical schemes.

\section{Complexity and discussion}
The given algorithm can be studied under two perspectives: as a hybrid quantum-classical algorithm or as a quantum-inspired classical numerical scheme. For the classicality of the streaming, in fact, this algorithm can be simulated fully on a classical computer. Nevertheless, we provide a computational analysis in both cases: (1) supposing the algorithm is run classically, as it was for the shown simulations , and (2) supposing that the state preparation, evolution and readout is made on a quantum computer.

\subsection{Classical}
When interpreted as a classical lattice method, the proposed QLG algorithm can be compared with LGCA and LBM. In this case, our algorithm has an asymptotic complexity of $O(NT)$ for simulating $N$ sites for $T$ time steps. Standard LGCA provides the cost of evolution plus the cost of averaging. Averaging, in fact, is necessary in standard LGCA. To have $N$ averaged sites we need to simulate $\tilde{N}$ sites with $\tilde{N}=\alpha N$ where $\alpha$ is the average step. Making $M$ ensemble simulations, we have a total cost of $O((M\alpha NT)$. So, accounting also for the averaging and ensemble costs, the QLG presented is more efficient than standard LGCA.

However, averaging is not strictly needed in LGCA. It can be avoided by calculating directly the mean populations, moving from a bit dynamics to an averaged-populations dynamics \cite{mcnamara2019use}. We can call this model ensemble LGCA. This lowers the cost to $O(NT)$, thus equating our algorithm. This is the same complexity of LBM, respect to which our algorithm does not achieve a computational advantage. While having the same complexity as LBM and ensemble LGCA, our algorithm can achieve better results. Ensemble LGCA, in fact, is unconditionally stable, but still has some unphysical features: non-Galilean invariance, artifical dependency of the pressure upon the velocity, non-tunable hydrodynamic parameters. Our model is also unconditionally stable, but does not show any non-Galilean invariance of the macroscopic equations, and has tunable hydrodynamic parameters. Moreover, it was proved that for some parameters the dynamics simulated by the present QLG cannot be calculated as the average of a lattice gas \cite{love2006type}, thus overcoming LGCA also on simulation possibilities. On the other side, LBM shows tunable parameters, but loses unconditional stability. To complete, entropic LBM have unconditional stability, and do not have any unphysicality, but they do not have tunable parameters. Our algorithm, instead, has both unconditional stability and tunable parameters, managing to achieve for enough long simulations also high Raynold's numbers. A complete comparison of the models is given in Table \ref{tab:classical}.

\begin{table*}[ht]
\caption{\label{tab:classical} Comparison as a classical numerical scheme, simulating $N$ sites for $T$ time steps}
\begin{ruledtabular}
\begin{tabular}{c c c c c c}
 %&\multicolumn{2}{c}{$D_{4h}^1$}&\multicolumn{2}{c}{$D_{4h}^5$}\\
 & Standard LGCA \cite{frisch2019lattice} & Ensemble LGCA \cite{mcnamara2019use} & SRT LBM \cite{kruger2016lattice} & Entropic LBM & This article \\ \hline
 Computational complexity & $O(M\alpha NT)$ & $O(NT)$ & $O(NT)$ & $O(NT)$ & $O(NT)$\\
 Stability\footnote{$\surd$ means unconditional stability, $\times$ means the lack of numerical stability in some regimes} & $\surd$ & $\surd$ & $\times$ & $\surd$ & $\surd$ \\
 Tunability\footnote{$\surd$ means tunable viscosity, $\times$ means viscosity cannot be tuned} & $\times$\footnote{Tunable in some specific cases \cite{boghosian2019cellular}} & $\times$ & $\surd$ & $\times$ & $\surd$\\
 Other & Unphysicalities\footnote{Non-Galilean invariance, pressure artifacts} & Unphysicalities\footnotemark[4] &  & \\
 %Ag& &$(4k)^{\text{a}}$& &$(4h)^{\text{a}}$\\
\end{tabular}
\end{ruledtabular}
\end{table*}

\subsection{Quantum}
When interpreted as an hybrid scheme, we consider the run on a quantum computer. The quantum computing architecture considered is a type-II quantum computer. This was implemented experimentally with nuclear magnetic resonance(NMR)\cite{pravia2003experimental}, and the model was simulated also for the diffusion equation \cite{berman2002simulation}. In this analysis, we do not consider any specific hardware, keeping a complexity analysis.

The quantum setup is straightforward: we have two qubits per cell, so the number of qubits $N_q$ for simulating $N$ cells is 
$$N_q=2N.$$
For the state preparation we apply one rotation $R_y(2\arcsin\sqrt{f_i})$ at each qubit. This gives a gate complexity of $O(1)$ per cell per time step. Then, we have the collision: in each cell we apply one two qubit unitary, resulting in a complexity of $O(1)$ as well. The streaming step is associated with the reinitialization step, which includes its cost. In a quantum setup, the readout is the important part. We need to estimate 2 observables per cell: $\hat{n}_0$ and $\hat{n}_1$. To do this, we need, considering a readout error of $\epsilon$, to do $O(1/\epsilon^2)$ samples in order to approximate well the postcollision $f_i$. It could be possible to apply specific amplitude amplification techniques, reducing the number of samples to $O(1/\epsilon)$ \cite{montanaro2015quantum}, but increasing the total gatecount.

Considering to have access to $N_q$ qubits, these evolve as independent pairs. This means that the initialization, collision and readout for different sites can be carried out in parallel. Also classically this can take place, thus confirming the lack of a quantum advantage in the present setup. This parallel quantum computation possibility brings the time complexity to be $O(T/\epsilon^2)$ for the evolution, including: initialization (streaming), collision and readout at each time step.

We can compare this cost with quantum nonlinear solvers (QNS) outlined in the introduction \cite{gan2025provably,bharadwaj2025quantum}, but with some attention. These algorithms, in fact, accomplish a different task respect to our algorithm: they produce a final quantum state whose amplitudes are proportional to the desired solution. They do this generally with $O(\log(N))$ qubits, showing a time complexity of $O(T,\log(N),h(\epsilon))$ where $h$ is the cost related to the accuracy of the algorithm $\epsilon$, possibly polynomial or polylogarithmic in $1/\epsilon$. The linearization technique affects the attainable nonlinear regimes, eventually limiting the possible viscosities in the case for Burger's equation. The concrete advantage depends on the task to complete. If the task regards the estimation of observables on the final state, QNS hints an advantage (to be confirmed on fault tolerant quantum computers). However, if the task is to retrieve the whole quantum state, then the comlexity needs to consider full state tomography, which will generally have a cost of $O(N)$, making QNS possibly less efficient than our algorithm, and other classical algorithms. This should come with no surprise, since it is clear from the first linear system solvers \cite{harrow2009quantum} that the advantage stands in the estimation of observables on a final quantum state. So, our algorithm is more efficient than the QNS examined in retrieving the complete final state of interest. Levareging on parallel quantum computation, the independent evolution of each cell guarantees that the cost is independent on the system size, which is not possible with full quantum state tomography of QNS.

Other quantum lattice Boltzmann schemes \cite{budinski2021quantum, wawrzyniak2025linearized} are promising when dealing with linear differential equations, even guaranteeing a quantum advantage comparable to quantum linear solvers employing Hamiltonian simulations. However, when dealing with nonlinearities these algorithms require our same tomography-reinitialization routine each time step, with the additional disadvantage that the routine cannot be in parallel as in our case. This proves that the better efficiency of our algorithms respect to other QLBM schemes for nonlinearities, as always expected from the difficulty in full-state retrieval. A summary of these comparisons is reported in Table \ref{tab:quantum}.

\begin{table*}
    \label{tab:quantum}
    \caption{Comparison as a quantum numerical scheme, simulating $N$ sites for $T$ time steps}
    \includegraphics[scale=1]{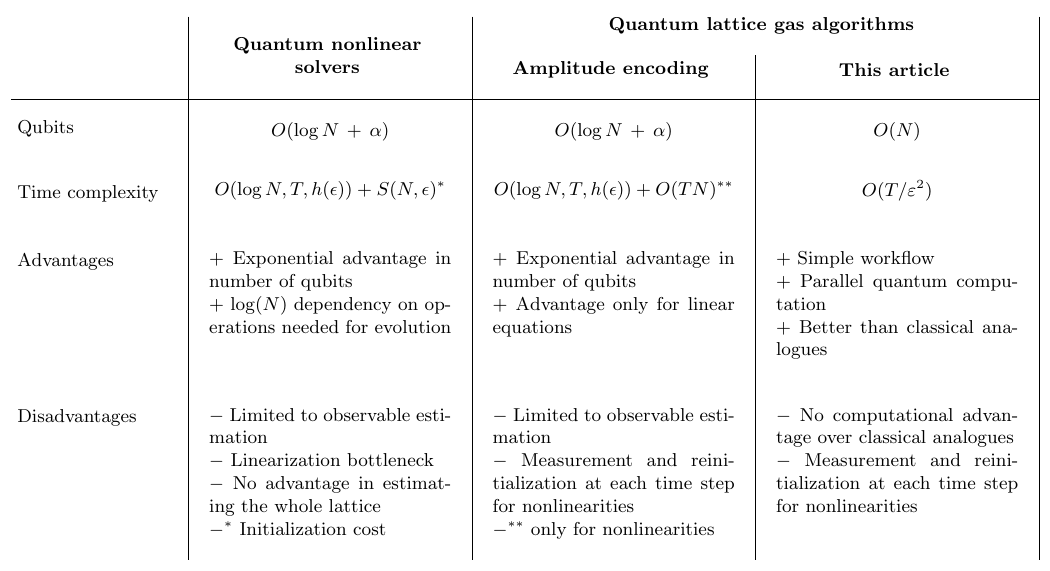}
\end{table*}

\section{Conclusion and perspectives}
In this work, we have analyzed the one-dimensional quantum lattice gas model proposed in \cite{yepez2006open}. In particular, we derived an analytical correction to the viscosity, which extends the validity of the Burgers-like equation to order $O(\epsilon^4)$ in the low-Mach regime. We discuss in detail the implications of this regime and provide numerical evidence supporting the correction, including comparisons with analytical solutions of the Burgers equation.

Furthermore, we highlighted the capability of this model to reproduce arbitrarily small viscosity for the Burger's equation, paying the price of longer simulation times or less grid points. 

In the second part, we proposed a minimal 2D extension of the algorithm. This model retains only two discrete velocities, preserving the analytical tractability and conceptual clarity of the 1D case. We demonstrated that, under these conditions, the scheme exhibits two-dimensional anisotropic Burgers-like behavior and derived the explicit form of the corresponding partial differential equation as a function of the chosen velocity set. 

The further interest in the newly developed 2D model arises from certain quantum walks. Specifically, quantum walks on a triangular lattice — with amplitudes defined on edges — which can reproduce 2D transport equations \cite{arrighi2018dirac,di2024quantum}. This suggests a potential path toward a quantum lattice gas model distinct from FHP, relying on fewer qubits while still capturing Navier-Stokes hydrodynamics. The reduced number of collisional qubits would also be better for simulations on real hardware.

Our algorithm obtains the complete final state for a nonlinear (Burger's) evolution of an initial function on $N$ points with $N_q=2N$ qubits, giving a time complexity of $O(T/\epsilon^2)$. This considers parallel quntum computation. In this task, it is more efficient than other nonlinear quantum solvers, and as efficient as other classical algorithms. Respect to these, however, it has both unconditional stability and tunable parameters, thus resulting as a promising alternative to LBM.

In conclusion, this algorithm explores the intersection between quantum computing and the simulation of nonlinearities, while using a simple workflow. In this, we achieve better computational features respect to other classical models, while having more efficient complete state retrieval respect to other quantum nonlinear solvers, as generally expected from classical algorithms. The structure of the algorithm is revisited as an efficient classical method, as well as a method to explore the quantum simulations of nonlinearities. Future developments include: introducing a third velocity to incorporate momentum conservation for approaching Navier–Stokes dynamics; studying possible isotropic alternatives with quantum walk-inspired streaming schemes; exploring the full quantum version of the algorithm, adopting a quantum streaming; assess the model’s performance on near-term quantum hardware. Such extensions would clarify whether this quantum lattice gas scheme can provide not only an efficient classical alternative but also a platform for realizing quantum advantages in computational fluid dynamics or quantum simulations.

%In conclusion, this model is promising for CFD simulations on quantum computers, since it combines analytical simplicity with tunable hydrodynamic parameters. Its unitary structure makes it a candidate framework for implementing quantum-native fluid simulations, potentially bridging the gap between quantum algorithms and classical lattice Boltzmann schemes. 

\begin{acknowledgments}
We are sincerely thankful to Jeffrey Yepez for the insights about the model, and for helpful discussions.

This work was partly supported by the PEPR EPiQ ANR-22-PETQ-0007, and ANR JCJC
DisQC ANR-22-CE47-0002-01

\end{acknowledgments}

\bibliographystyle{apsrev4-2}
\bibliography{apssamp}% Produces the bibliography via BibTeX.

@book{wolf2004lattice,
  title={Lattice-gas cellular automata and lattice Boltzmann models: an introduction},
  author={Wolf-Gladrow, Dieter A},
  year={2004},
  publisher={Springer}
}

@incollection{mcnamara2019use,
  title={Use of the Boltzmann equation to simulate lattice-gas automata},
  author={McNamara, Guy G and Zanetti, Gianluigi},
  booktitle={Lattice Gas Methods For Partial Differential Equations},
  pages={289--296},
  year={2019},
  publisher={CRC Press}
}

@incollection{boghosian2019cellular,
  title={A cellular automaton for Burgers' equation},
  author={Boghosian, Bruce M and Levermore, C David},
  booktitle={Lattice Gas Methods For Partial Differential Equations},
  pages={483--496},
  year={2019},
  publisher={CRC Press}
}

@article{yepez2006open,
  title={Open quantum system model of the one-dimensional Burgers equation with tunable shear viscosity},
  author={Yepez, Jeffrey},
  journal={Physical Review A—Atomic, Molecular, and Optical Physics},
  volume={74},
  number={4},
  pages={042322},
  year={2006},
  publisher={APS}
}

@article{micci2015measurement,
  title={Measurement-based quantum lattice gas model of fluid dynamics in 2+ 1 dimensions},
  author={Micci, Michael M and Yepez, Jeffrey},
  journal={Physical Review E},
  volume={92},
  number={3},
  pages={033302},
  year={2015},
  publisher={APS}
}

@article{pravia2003experimental,
  title={Experimental demonstration of quantum lattice gas computation},
  author={Pravia, Marco A and Chen, Zhiying and Yepez, Jeffrey and Cory, David G},
  journal={Quantum Information Processing},
  volume={2},
  number={1},
  pages={97--116},
  year={2003},
  publisher={Springer}
}

@article{montanaro2015quantum,
  title={Quantum speedup of Monte Carlo methods},
  author={Montanaro, Ashley},
  journal={Proceedings. Mathematical, Physical, and Engineering Sciences/The Royal Society},
  volume={471},
  number={2181},
  pages={20150301},
  year={2015}
}

@article{berman2002simulation,
  title={Simulation of the diffusion equation on a type-II quantum computer},
  author={Berman, GP and Ezhov, AA and Kamenev, DI and Yepez, J},
  journal={Physical Review A},
  volume={66},
  number={1},
  pages={012310},
  year={2002},
  publisher={APS}
}

@article{budinski2021quantum,
  title={Quantum algorithm for the advection--diffusion equation simulated with the lattice Boltzmann method},
  author={Budinski, Ljubomir},
  journal={Quantum Information Processing},
  volume={20},
  number={2},
  pages={57},
  year={2021},
  publisher={Springer}
}

@article{xiao2025quantum,
  title={Quantum Lattice Kinetic Scheme for Solving Two-dimensional and Three-dimensional Incompressible Flows},
  author={Xiao, Yang and Yang, Liming and Shu, Chang and Du, Yinjie and Dong, Hao and Wu, Jie},
  journal={arXiv preprint arXiv:2505.10883},
  year={2025}
}

@article{xu2025improved,
  title={Improved quantum lattice Boltzmann method for advection-diffusion equations with a linear collision model},
  author={Xu, Li and Li, Ming and Zhang, Lei and Sun, Hai and Yao, Jun},
  journal={Physical Review E},
  volume={111},
  number={4},
  pages={045305},
  year={2025},
  publisher={APS}
}

@article{wawrzyniak2024unitary,
  title={Unitary Quantum Algorithm for the Lattice-Boltzmann Method},
  author={Wawrzyniak, David and Winter, Josef and Schmidt, Steffen and Indinger, Thomas and Schramm, Uwe and Jan{\ss}en, Christian and Adams, Nikolaus A},
  journal={arXiv preprint arXiv:2405.13391},
  year={2024}
}

@inproceedings{bediche2025fully,
  title={Fully Quantum Algorithm for the 1-dimensional linear Lattice Boltzmann Method},
  author={Bediche, Mohammed and van Waveren, Matthijs and Ricot, Denis and Sagaut, Pierre},
  booktitle={QUEST IS 25},
  year={2025}
}

@article{fonio2025adaptive,
  title={Adaptive Lattice Gas Algorithm: Classical and Quantum implementations},
  author={Fonio, Niccol{\`o} and Budinski, Ljubomir and Lahtinen, Valtteri and Sagaut, Pierre},
  journal={arXiv preprint arXiv:2504.13549},
  year={2025}
}

@article{zamora2025float,
  title={Float Lattice Gas Automata: A connection between Molecular Dynamics and Lattice Boltzmann Method for quantum computers},
  author={Zamora, Antonio David Bastida and Budinski, Ljubomir and Lahtinen, Valtteri and Sagaut, Pierre},
  journal={arXiv preprint arXiv:2503.23750},
  year={2025}
}

@article{wang2025quantum,
  title={Quantum lattice Boltzmann method for simulating nonlinear fluid dynamics},
  author={Wang, Boyuan and Meng, Zhaoyuan and Zhao, Yaomin and Yang, Yue},
  journal={npj Quantum Information},
  year={2025},
  publisher={Nature Publishing Group UK London}
}

@book{succi2001lattice,
  title={The lattice Boltzmann equation: for fluid dynamics and beyond},
  author={Succi, Sauro},
  year={2001},
  publisher={Oxford university press}
}

@article{sanavio2024lattice,
  title={Lattice Boltzmann--Carleman quantum algorithm and circuit for fluid flows at moderate Reynolds number},
  author={Sanavio, Claudio and Succi, Sauro},
  journal={AVS Quantum Science},
  volume={6},
  number={2},
  year={2024},
  publisher={AIP Publishing}
}

@article{zecchi2025improved,
  title={Improved amplitude amplification strategies for the quantum simulation of classical transport problems},
  author={Zecchi, Alessandro Andrea and Sanavio, Claudio and Perotto, Simona and Succi, Sauro},
  journal={Quantum Science and Technology},
  volume={10},
  number={3},
  pages={035039},
  year={2025},
  publisher={IOP Publishing}
}

@article{yepez2009vortex,
  title={Vortex-antivortex pair in a Bose-Einstein condensate},
  author={Yepez, Jeffrey and Vahala, George and Vahala, Linda},
  journal={The European Physical Journal-Special Topics},
  volume={171},
  number={1},
  pages={9--14},
  year={2009},
  publisher={Springer}
}

@article{yepez2009superfluid,
  title={Superfluid turbulence from quantum Kelvin wave to classical Kolmogorov cascades},
  author={Yepez, Jeffrey and Vahala, George and Vahala, Linda and Soe, Min},
  journal={Physical Review Letters},
  volume={103},
  number={8},
  pages={084501},
  year={2009},
  publisher={APS}
}

@article{ljubomir2022quantum,
  title={Quantum algorithm for the Navier--Stokes equations by using the streamfunction-vorticity formulation and the lattice Boltzmann method},
  author={Ljubomir, Budinski},
  journal={International Journal of Quantum Information},
  volume={20},
  number={02},
  pages={2150039},
  year={2022},
  publisher={World Scientific}
}

@article{kumar2025quantum,
  title={Quantum unitary matrix representation of the lattice Boltzmann model for low Reynolds fluid flow simulation},
  author={Kumar, E Dinesh and Frankel, Steven H},
  journal={AVS Quantum Science},
  volume={7},
  number={1},
  year={2025},
  publisher={AIP Publishing}
}

@article{lewis2024limitations,
  title={Limitations for quantum algorithms to solve turbulent and chaotic systems},
  author={Lewis, Dylan and Eidenbenz, Stephan and Nadiga, Balasubramanya and Suba{\c{s}}{\i}, Yi{\u{g}}it},
  journal={Quantum},
  volume={8},
  pages={1509},
  year={2024},
  publisher={Verein zur F{\"o}rderung des Open Access Publizierens in den Quantenwissenschaften}
}

@article{love2006type,
  title={Type II quantum algorithms},
  author={Love, Peter J and Boghosian, Bruce M},
  journal={Physica A: Statistical Mechanics and its Applications},
  volume={362},
  number={1},
  pages={210--214},
  year={2006},
  publisher={Elsevier}
}

@article{love2019quantum,
  title={On quantum extensions of hydrodynamic lattice gas automata},
  author={Love, Peter},
  journal={Condensed Matter},
  volume={4},
  number={2},
  pages={48},
  year={2019},
  publisher={MDPI}
}

@incollection{frisch2019lattice,
  title={Lattice gas hydrodynamics in two and three dimensions},
  author={Frisch, Uriel and d'Humieres, Dominique and Hasslacher, Brosl and Lallemand, Pierre and Pomeau, Yves and Rivet, Jean-Pierre},
  booktitle={Lattice Gas Methods for Partial Differential Equations},
  pages={77--136},
  year={2019},
  publisher={CRC Press}
}

@book{kruger2016lattice,
  title={Lattice Boltzmann Method-Principles and Practice},
  author={Kruger, Timm},
  year={2016},
  publisher={Springer International Publish}
}

@article{succi2023quantum,
  title={Quantum computing for fluids: Where do we stand?},
  author={Succi, Sauro and Itani, Wael and Sreenivasan, Katepalli and Steijl, Ren{\'e}},
  journal={Europhysics Letters},
  volume={144},
  number={1},
  pages={10001},
  year={2023},
  publisher={IOP Publishing}
}

@book{di2024quantum,
  title={Quantum Walks, Limits, and Transport Equations},
  author={Di Molfetta, Giuseppe},
  year={2024},
  publisher={IOP Publishing}
}

@article{arrighi2018dirac,
  title={Dirac equation as a quantum walk over the honeycomb and triangular lattices},
  author={Arrighi, Pablo and Di Molfetta, Giuseppe and M{\'a}rquez-Mart{\'\i}n, Iv{\'a}n and P{\'e}rez, Armando},
  journal={Physical Review A},
  volume={97},
  number={6},
  pages={062111},
  year={2018},
  publisher={APS}
}

@article{wissocq2019extended,
  title={An extended spectral analysis of the lattice Boltzmann method: modal interactions and stability issues},
  author={Wissocq, Gauthier and Sagaut, Pierre and Boussuge, Jean-Fran{\c{c}}ois},
  journal={Journal of Computational Physics},
  volume={380},
  pages={311--333},
  year={2019},
  publisher={Elsevier}
}

@article{wissocq2022hydrodynamic,
  title={Hydrodynamic limits and numerical errors of isothermal lattice Boltzmann schemes},
  author={Wissocq, Gauthier and Sagaut, Pierre},
  journal={Journal of Computational Physics},
  volume={450},
  pages={110858},
  year={2022},
  publisher={Elsevier}
}

@article{hosseini2023entropic,
  title={Entropic lattice Boltzmann methods: A review},
  author={Hosseini, Seyed Ali and Atif, Mohammad and Ansumali, Santosh and Karlin, Iliya V},
  journal={Computers \& Fluids},
  volume={259},
  pages={105884},
  year={2023},
  publisher={Elsevier}
}

@inproceedings{berry2014exponential,
  title={Exponential improvement in precision for simulating sparse Hamiltonians},
  author={Berry, Dominic W and Childs, Andrew M and Cleve, Richard and Kothari, Robin and Somma, Rolando D},
  booktitle={Proceedings of the forty-sixth annual ACM symposium on Theory of computing},
  pages={283--292},
  year={2014}
}

@article{low2019hamiltonian,
  title={Hamiltonian simulation by qubitization},
  author={Low, Guang Hao and Chuang, Isaac L},
  journal={Quantum},
  volume={3},
  pages={163},
  year={2019},
  publisher={Verein zur F{\"o}rderung des Open Access Publizierens in den Quantenwissenschaften}
}

@inproceedings{gilyen2019quantum,
  title={Quantum singular value transformation and beyond: exponential improvements for quantum matrix arithmetics},
  author={Gily{\'e}n, Andr{\'a}s and Su, Yuan and Low, Guang Hao and Wiebe, Nathan},
  booktitle={Proceedings of the 51st annual ACM SIGACT symposium on theory of computing},
  pages={193--204},
  year={2019}
}

@article{berry2017quantum,
  title={Quantum algorithm for linear differential equations with exponentially improved dependence on precision},
  author={Berry, Dominic W and Childs, Andrew M and Ostrander, Aaron and Wang, Guoming},
  journal={Communications in Mathematical Physics},
  volume={356},
  number={3},
  pages={1057--1081},
  year={2017},
  publisher={Springer}
}

@article{joseph2020koopman,
  title={Koopman--von Neumann approach to quantum simulation of nonlinear classical dynamics},
  author={Joseph, Ilon},
  journal={Physical Review Research},
  volume={2},
  number={4},
  pages={043102},
  year={2020},
  publisher={APS}
}

@article{gan2025provably,
  title={Provably Efficient Quantum Algorithms for Solving Nonlinear Differential Equations Using Multiple Bosonic Modes Coupled with Qubits},
  author={Gan, Yu and Alipanah, Hirad and Cheng, Jinglei and Wu, Zeguan and Li, Guangyi and Mendoza-Arenas, Juan Jos{\'e} and Givi, Peyman and Malik, Mujeeb R and McDermott, Brian J and Liu, Junyu},
  journal={arXiv preprint arXiv:2511.09939},
  year={2025}
}

@article{bharadwaj2025quantum,
  title={Quantum Homotopy Algorithm for Solving Nonlinear PDEs and Flow Problems},
  author={Bharadwaj, Sachin S and Nadiga, Balasubramanya and Eidenbenz, Stephan and Sreenivasan, Katepalli R},
  journal={arXiv preprint arXiv:2512.21033},
  year={2025}
}

@article{bharadwaj2025compact,
  title={Compact quantum algorithms for time-dependent differential equations},
  author={Bharadwaj, Sachin S and Sreenivasan, Katepalli R},
  journal={Physical Review Research},
  volume={7},
  number={2},
  pages={023262},
  year={2025},
  publisher={APS}
}

@article{jin2022quantum,
  title={Quantum simulation of partial differential equations via Schrodingerisation},
  author={Jin, Shi and Liu, Nana and Yu, Yue},
  journal={arXiv preprint arXiv:2212.13969},
  year={2022}
}

@article{hu2024quantum,
  title={Quantum circuits for partial differential equations via Schr{\"o}dingerisation},
  author={Hu, Junpeng and Jin, Shi and Liu, Nana and Zhang, Lei},
  journal={Quantum},
  volume={8},
  pages={1563},
  year={2024},
  publisher={Verein zur F{\"o}rderung des Open Access Publizierens in den Quantenwissenschaften}
}

@article{liu2021efficient,
  title={Efficient quantum algorithm for dissipative nonlinear differential equations},
  author={Liu, Jin-Peng and Kolden, Herman {\O}ie and Krovi, Hari K and Loureiro, Nuno F and Trivisa, Konstantina and Childs, Andrew M},
  journal={Proceedings of the National Academy of Sciences},
  volume={118},
  number={35},
  pages={e2026805118},
  year={2021},
  publisher={National Academy of Sciences}
}

@article{li2025potential,
  title={Potential quantum advantage for simulation of fluid dynamics},
  author={Li, Xiangyu and Yin, Xiaolong and Wiebe, Nathan and Chun, Jaehun and Schenter, Gregory K and Cheung, Margaret S and M{\"u}lmenst{\"a}dt, Johannes},
  journal={Physical Review Research},
  volume={7},
  number={1},
  pages={013036},
  year={2025},
  publisher={APS}
}

@article{jennings2025end,
  title={An end-to-end quantum algorithm for nonlinear fluid dynamics with bounded quantum advantage},
  author={Jennings, David and Korzekwa, Kamil and Lostaglio, Matteo and Ashworth, Richard and Marsili, Emanuele and Rolston, Stephen},
  journal={arXiv preprint arXiv:2512.03758},
  year={2025}
}

@article{itani2024quantum,
  title={Quantum algorithm for lattice Boltzmann (QALB) simulation of incompressible fluids with a nonlinear collision term},
  author={Itani, Wael and Sreenivasan, Katepalli R and Succi, Sauro},
  journal={Physics of Fluids},
  volume={36},
  number={1},
  year={2024},
  publisher={AIP Publishing}
}

@article{jennings2026quantum,
  title={Quantum Koopman Algorithms},
  author={Jennings, David and Korzekwa, Kamil and Lostaglio, Matteo and Wang, Guoming},
  journal={arXiv preprint arXiv:2605.19054},
  year={2026}
}

@article{gonzalez2025quantum,
  title={Quantum Carleman linearization efficiency in nonlinear fluid dynamics},
  author={Gonzalez-Conde, Javier and Lewis, Dylan and Bharadwaj, Sachin S and Sanz, Mikel},
  journal={Physical Review Research},
  volume={7},
  number={2},
  pages={023254},
  year={2025},
  publisher={APS}
}

@article{harrow2009quantum,
  title={Quantum algorithm for linear systems of equations},
  author={Harrow, Aram W and Hassidim, Avinatan and Lloyd, Seth},
  journal={Physical review letters},
  volume={103},
  number={15},
  pages={150502},
  year={2009},
  publisher={APS}
}

@article{wawrzyniak2025linearized,
  title={Linearized quantum lattice-Boltzmann method for the advection-diffusion equation using dynamic circuits},
  author={Wawrzyniak, David and Winter, Josef and Schmidt, Steffen and Indinger, Thomas and Jan{\ss}en, Christian F and Schramm, Uwe and Adams, Nikolaus A},
  journal={Computer Physics Communications},
  pages={109856},
  year={2025},
  publisher={Elsevier}
}

@misc{foniocode,
  author = {N. Fonio},
  title = {Github repository},
  howpublished = {\url{https://github.com/niccolofonio/QuantumLatticeBoltzmannGas}},
  year={2026}
}

\clearpage

\appendix

\section{Full 1D calculations}
We start from finite difference expression

\begin{equation} \label{appendix_eq: finite_diff_1d}
    f_i(x-c_i \delta x,t + \delta t) = f_i(x,t) + \Omega_i(x,t).
\end{equation}

We know that $\Omega_i=c_i\Omega$ with
\begin{equation} \label{appendix_eq:omega}
    \begin{split}
        \Omega & = (f_0-f_1)\sin^2\theta + \\ 
        & + \sin(2\theta)\cos(\zeta-\xi)\sqrt{f_0(1-f_0)f_1(1-f_1)}.
    \end{split}
\end{equation}

If we solve $\Omega=0$, which is done in \cite{yepez2006open}, it is possible to derive the explicit equilibrium populations
\small
\begin{equation} \label{appendix_eq:equilibria}
    f_i^{eq}=\frac{\rho}{2} +\frac{c_i}{2\alpha} \qty[\sqrt{\alpha^2+1-2\alpha^2\rho+\alpha^2\rho^2}-\sqrt{\alpha^2+1}].
\end{equation}
\normalsize
These bring correctly to $\rho=f_1^{eq}+f_0^{eq}$. We will also explicitly use the momentum $u=f_1^{eq}-f_0^{eq}$, explicitly

\begin{equation} \label{appendix_eq:u}
    u = -\frac{1}{\alpha} \qty(\sqrt{1+\alpha^2} - \sqrt{1+\alpha^2(\rho-1)^2}).
\end{equation}

We take now Eq. \ref{appendix_eq: finite_diff_1d} and we do a Taylor expansion. We keep in mind that later we will do a Chapman-Anskog expansion, thus we will have $\delta x\approx \epsilon$ and $\delta t\approx \epsilon^2$, where $\epsilon$ is the Knudsen number. We obtain wit the Taylor expansion

\begin{equation} \label{appendix_eq:taylor1}
    -c_i \delta x \de_xf_i+\delta t \de_tf_i+\frac{\delta x^2}{2}c_i^2\de_{xx}f_i=c_i \Omega.
\end{equation}
Now we do the Chapman-Enskog expansion, that is considering
\begin{equation}
    f_i\approx f_i^{eq} + \epsilon f_i^{(1)} + \epsilon^2 f_i^{(2)} + O(\epsilon^3).
\end{equation}
Thus we obtain
\begin{widetext}
\begin{equation} \label{appendix_eq:taylor2}
\begin{split}
-c_i\dx \de_xf_i^{eq} - c_i\dx \epsilon \de_xf_i^{(1)} +\dt\de_tf_i^{eq}+\frac{\dx^2}{2}c_i^2\de_{xx}f_i^{eq} + O(\epsilon^3) = c_i \Omega[f_i^{eq}] + \epsilon c_i[J_0f_0^{(1)} + J_1f_1^{(1)}],
\end{split}
\end{equation}
\end{widetext}
with $J_i=\pdv{\Omega}{f_i}\big|_{eq}$. We know that $\Omega[f_i^{eq}]=0$. We can focus then on the first order, where we have

\begin{equation}
    -c_i\dx\de_xf_i^{eq}=\epsilon c_i[J_0f_0^{(1)} + J_1f_1^{(1)}].
\end{equation}
Matrix notation as used in \cite{yepez2006open} is very convenient. However, we can alternatively simplify $c_i$ and $\epsilon$ and use some algebraic manipulations to obtain

\begin{equation}
    J_0f_0^{(1)} + J_1f_1^{(1)} = \frac{1}{2}(f_1^{(1)}-f_0^{(1)})(J_1-J_0).
\end{equation}
From this it is straightforward to have

\begin{equation} \label{appendix_eq:deltaf1}
    \epsilon (f_1^{(1)}-f_0^{(1)}) = -\frac{1}{J_1-J_0}\de_x \rho.
\end{equation}
This is the crucial expression that needs to be calculated explicitly to get the correction to the viscosity. From trigonometric identities we have
\begin{equation}
    \sin(2\theta)\cos(\zeta-\xi) = 2\alpha\sin^2(\theta),
\end{equation} 
knowing $\alpha=\cot(\theta)\cos(\zeta-\xi)$. We have in particular, using $\bar{f_i}=1-f_i$
\small
\begin{align}
    J_0 & = \sin^2(\theta) \qty[1+ 2\alpha\qty(\frac{\sqrt{f_0\bar{f_0}f_1}}{2\sqrt{f_0}}-\frac{\sqrt{f_0f_1\bar{f_1}}}{2\sqrt{\bar{f_0}}}) ], \\
    J_1 & = \sin^2(\theta) \qty[-1+ 2\alpha\qty(\frac{\sqrt{f_0\bar{f_0}\bar{f_1}}}{2\sqrt{f_1}}-\frac{\sqrt{f_0\bar{f_0}f_1}}{2\sqrt{\bar{f_1}}})] .
\end{align}
\normalsize
We define for simplicity $a=\sqrt{f_0\bar{f_0}f_1\bar{f_1}}$ and calculate

\begin{align}
    J_1-J_0 & = -2\sin^2(\theta)+\alpha \sin^2(\theta)a\qty[\frac{1}{f_1} - \frac{1}{\bar{f_1}} - \frac{1}{f_0} + \frac{1}{\bar{f_0}}] \notag \\
    & = -2\sin^2(\theta)+\alpha \sin^2(\theta)a A.
\end{align}
It is possible to calculate
\begin{equation}
    A=\frac{(f_0-f_1)}{a^2}(1-f_0-f_1+2f_0f_1).
\end{equation}
Then, from the equilibrium condition it is possible to see that
\begin{equation}
    \frac{f_0-f_1}{a} = -2\alpha.
\end{equation}
Thus using the definitions of $\rho$ and $u$, we have
\small
\begin{equation}
    J_1-J_0 = -2\sin^2(\theta)-2\alpha^2\sin^2(\theta) \qty[1-\rho+\frac{\rho^2-u^2}{2}].
\end{equation}
\normalsize
This is consistent with Yepez result, but here we managed to simplify its explicit expression, which will be needed to calculate the correction to the viscosity. It is possible to calculate this relation
\small
\begin{equation}
\begin{split}
    \frac{\rho^2-u^2}{2} = & -\frac{1}{\alpha^2} + (\rho-1)+ \\
    & + \frac{1}{\alpha^2} \qty[ \sqrt{\alpha^2+1} \sqrt{\alpha^2+1-2\alpha^2\rho+\alpha^2\rho^2} ].
\end{split}
\end{equation}
\normalsize
Thus, the final expression that we aimed for is
\small
\begin{equation} \label{appendix_eq:finalDJ}
\begin{split}
    J_1-J_0 = & -2\sin^2(\theta)  \bigg[ \alpha^2(\rho-1)  \\
    & + \sqrt{\alpha^2+1} \sqrt{\alpha^2+1-2\alpha^2\rho+\alpha^2\rho^2} \bigg].
\end{split}
\end{equation}
\normalsize
For further notice we calculate now these derivatives 
%\begin{widetext}
\small
\begin{equation} \label{appendix_eq:dxDJ}
    \begin{split}
        \pdv{(J_1-J_0)}{x} & =-2\sin^2(\theta)\bigg[ \alpha^2\de_x\rho +\\ &+\frac{\sqrt{\alpha^2+1}}{2} \frac{-2\alpha^2\de_x\rho+2\alpha^2\rho\de_x\rho}{\sqrt{\alpha^2+1-2\alpha^2\rho+\alpha^2\rho^2}} \bigg],
    \end{split}
\end{equation}
\normalsize
\begin{equation}\label{appendix_eq:drhoDJ}
    \pdv{(J_1-J_0)}{\rho} = \sin^2(\theta) \frac{(2-2\rho)(\alpha^2+1)^{3/2}}{\sqrt{\alpha^2(\rho-1)^2+1}}. 
\end{equation}    
%\end{widetext}
Now we go back to Eq. \ref{appendix_eq:taylor2}. We need to calculate the sum of those expressions $(i=0)+(i=1)$. This is because we are conserving mass. Technically it would be possible to write this expression for each conserved quantity. The right-hand side (RHS) vanishes in this sum operation, since we had $\Omega_i=c_i\Omega$ and also for any further orders in $\epsilon$. For the left-hand side(LHS) we obtain

\begin{equation}
\begin{split}
    \dt \de_t\rho -& \dx\de_x(f_1^{eq} -f_0^{eq}) - \dx\epsilon \de_x (f_1^{(1)}-f_0^{(1)})+ \\ &  \frac{\dx^2}{2}\de_{xx}\rho + O(\epsilon^3) = 0.
    \end{split}
\end{equation}
Using Eq. \ref{appendix_eq:deltaf1}, we rewrite
\begin{widetext}
% \begin{equation}
%     \dt \de_t\rho - \dx\de_xu + \dx^2 \de_x \qty[\frac{1}{J_1-J_0}\de_x \rho]+\frac{\dx^2}{2}\de_{xx}\rho + O(\epsilon^3) = 0
% \end{equation}
\begin{equation}\label{appendix_eq:pde1}
    \dt \de_t\rho - \dx\de_xu - \dx^2 \de_x\rho \qty[\frac{\de_x(J_1-J_0)}{(J_1-J_0)^2}] + \dx^2 \frac{1}{J_1-J_0}\de_{xx} \rho + \frac{\dx^2}{2}\de_{xx}\rho + O(\epsilon^3) = 0.
\end{equation}
\end{widetext}
From this expression we need to calculate the low-Mach approximation, that consists in $\rho\approx 1$. Thus we need to expand the different expressions around $\rho = 1$

Defining $g(\rho)=1/(J_1-J_0)$ we have
\small
\begin{align}
    g(\rho) & \approx g(\rho=1)+\pdv{g}{\rho}\bigg|_{\rho=1} (\rho-1) + O((\rho-1)^2) \\
    & \approx -\frac{1}{2\sin^2(\theta)}\frac{1}{\sqrt{\alpha^2+1}} + \frac{\de_\rho (J_1-J_0)}{(J_1-J_0)^2}\bigg|_{\rho=1} (\rho-1).
\end{align}
\normalsize
We know from Eq. \ref{appendix_eq:drhoDJ} that
$$\frac{\de_\rho (J_1-J_0)}{(J_1-J_0)^2}\bigg|_{\rho=1}=0.$$
Thus we obtain, supposing $O( (\rho-1))=O(\epsilon)$,
\begin{equation}
    \frac{1}{J_1-J_0}\approx -\frac{1}{2\sin^2(\theta)}\frac{1}{\sqrt{\alpha^2+1}} + O( \epsilon^2).
\end{equation}
We can notice a term that is 
\begin{equation}
    \epsilon\dx\de_x\rho \frac{\de_x(J_1-J_0)}{(J_1-J_0)^2}.
\end{equation}
From Eq. \ref{appendix_eq:dxDJ} we have that $\de_x(J_1-J_0)\propto \de_x \rho$. Considering that in the low-Mach regime $\de_x\rho\approx O(\epsilon)$, it is clear to see that this term can be neglected. Last but not least we consider the term $\de_x u$.
\begin{equation}
    \de_x u = \frac{\alpha(\rho-1)\de_x\rho}{\sqrt{1+\alpha^2(\rho-1)^2}}.
\end{equation}
Now again we define $h(\rho)=\de_x u$ and calculate the expansion around $\rho=1$. We keep in mind, however, that we will need to go one order further, because this term is at order $\dx$. Thus we have
\small
\begin{equation}
    h(\rho)\approx h(\rho=1) + \pdv{h}{\rho} \bigg|_{\rho=1}(\rho-1)+\frac{1}{2}\pdv[2]{h}{\rho} \bigg|_{\rho=1} (\rho-1)^2
\end{equation}
\normalsize
We can calculate these terms, having
\begin{align}
    h(\rho=1) & = 0,  \\
    \de_{\rho}\de_x \rho \big|_{\rho=1} & = 0, \\
    \pdv{h}{\rho} = \frac{\alpha}{(\alpha^2(\rho-1)^2+1)^{3/2}}\bigg|_{\rho=1} &=\alpha, \\
    \pdv[2]{h}{\rho} = \frac{6\alpha^3(\rho-1)}{(\alpha^2(\rho-1)^2+1)^{5/2}}\bigg|_{\rho=1} &=0.
\end{align}
Thus, we finally get
\begin{equation}
    \de_x u \approx \alpha(\rho-1) + O(\epsilon^3).
\end{equation}
We can finally write Eq. \ref{appendix_eq:pde1} in the low-Mach regime as
\begin{widetext}
\begin{equation}
    \de_t\rho+c\alpha(1-\rho)\de_x\rho+\frac{\dx^2}{\dt}\frac{1}{2} \qty(1-\frac{1}{\sin^2(\theta)\sqrt{\alpha^2+1}})\de_{xx}\rho + O(\epsilon^4)= 0.
\end{equation}
\end{widetext}
%The final expression is reported also in Eq.\ref{eq:final_pde_1d_bis}

\begin{figure*}
    \centering
    \includegraphics[width=0.23\linewidth]{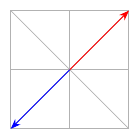}
    \includegraphics[width=0.76\linewidth]{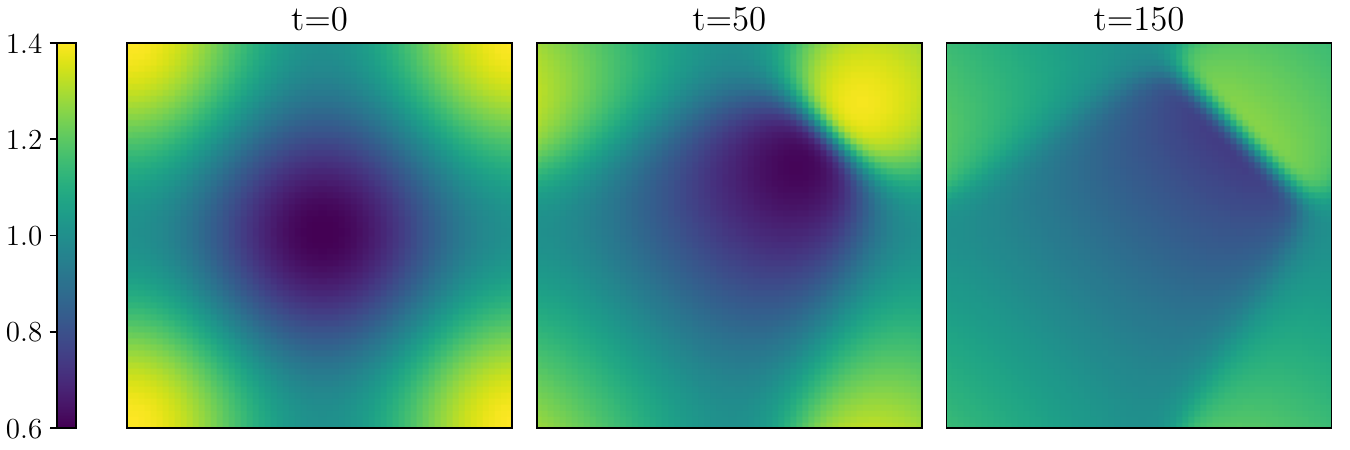}
    \includegraphics[width=0.23\linewidth]{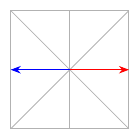}
    \includegraphics[width=0.76\linewidth]{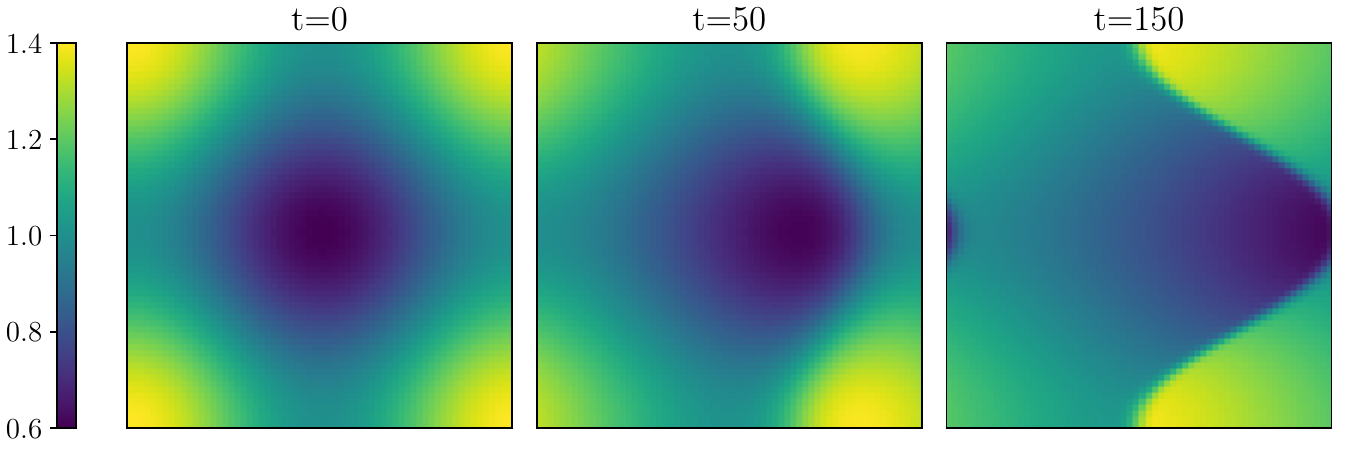}
    \includegraphics[width=0.23\linewidth]{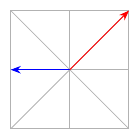}
    \includegraphics[width=0.76\linewidth]{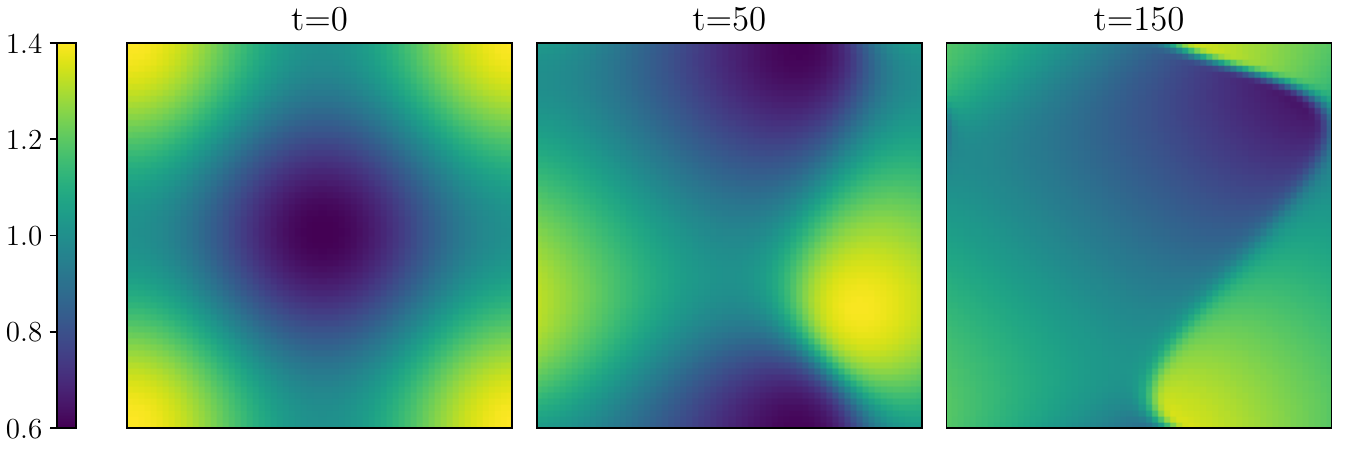}
    \includegraphics[width=0.23\linewidth]{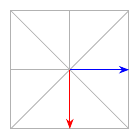}
    \includegraphics[width=0.76\linewidth]{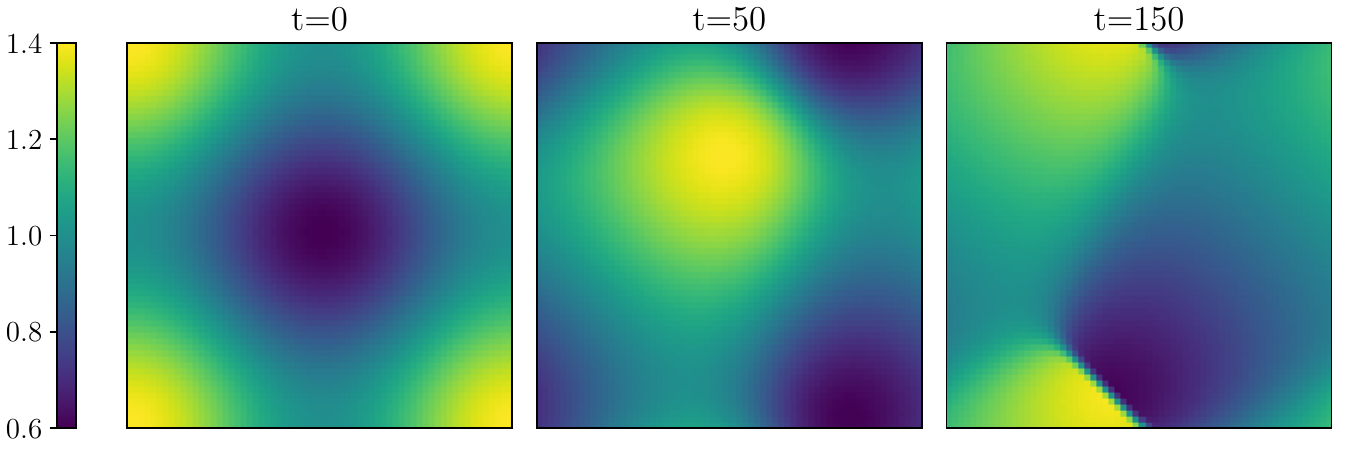}
    \caption{Simulations of a 64$\times$64 square grid for different discrete velocity sets. The blue (red) arrow represents the streaming direction of $f_0$ ($f_1$). The lattice is initialized as in Eq. \ref{eq:initial_cos}}
    \label{fig:2D_simulations}
\end{figure*}
\begin{figure*}
    \centering
    \includegraphics[width=0.24\linewidth]{fig_9_10_arrows_set1.pdf}
    \includegraphics[width=0.7\linewidth]{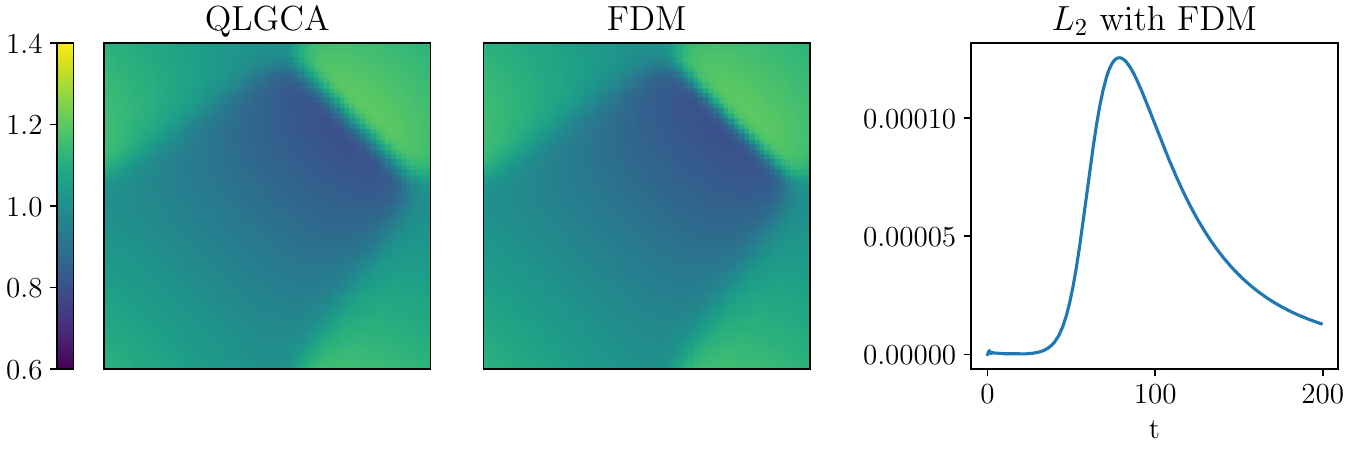}
    \includegraphics[width=0.24\linewidth]{fig_9_10_arrows_set2.pdf}
    \includegraphics[width=0.7\linewidth]{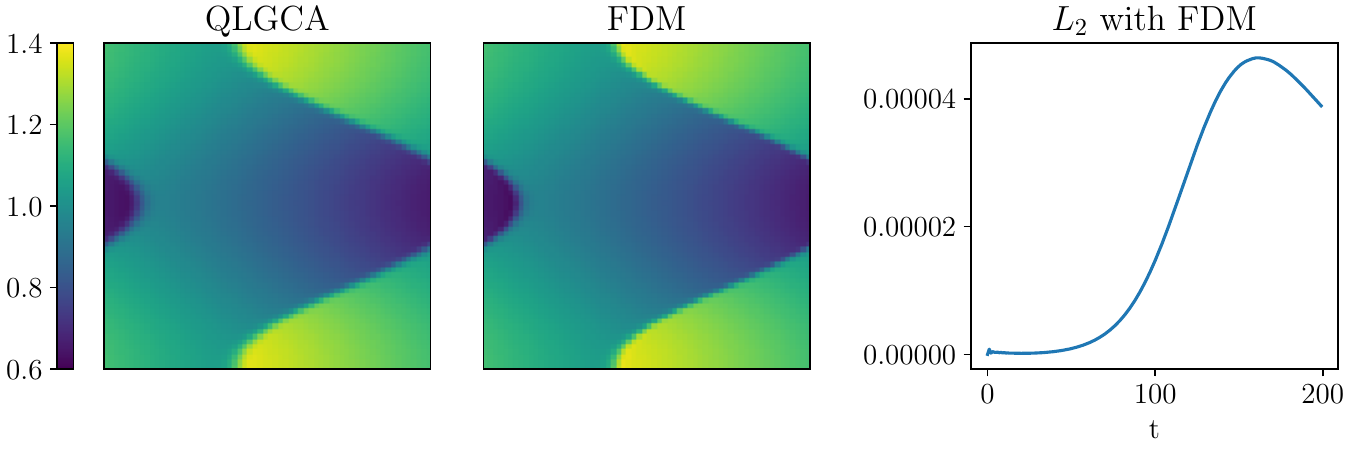}
    \includegraphics[width=0.24\linewidth]{fig_9_10_arrows_set3.pdf}
    \includegraphics[width=0.7\linewidth]{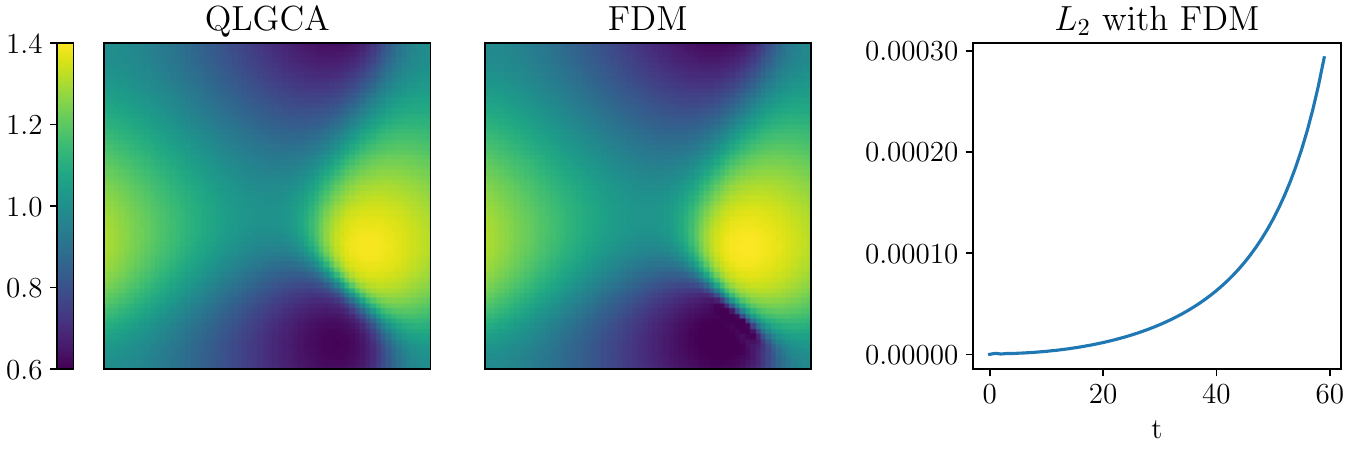}
    \includegraphics[width=0.24\linewidth]{fig_9_10_arrows_set4.pdf}
    \includegraphics[width=0.7\linewidth]{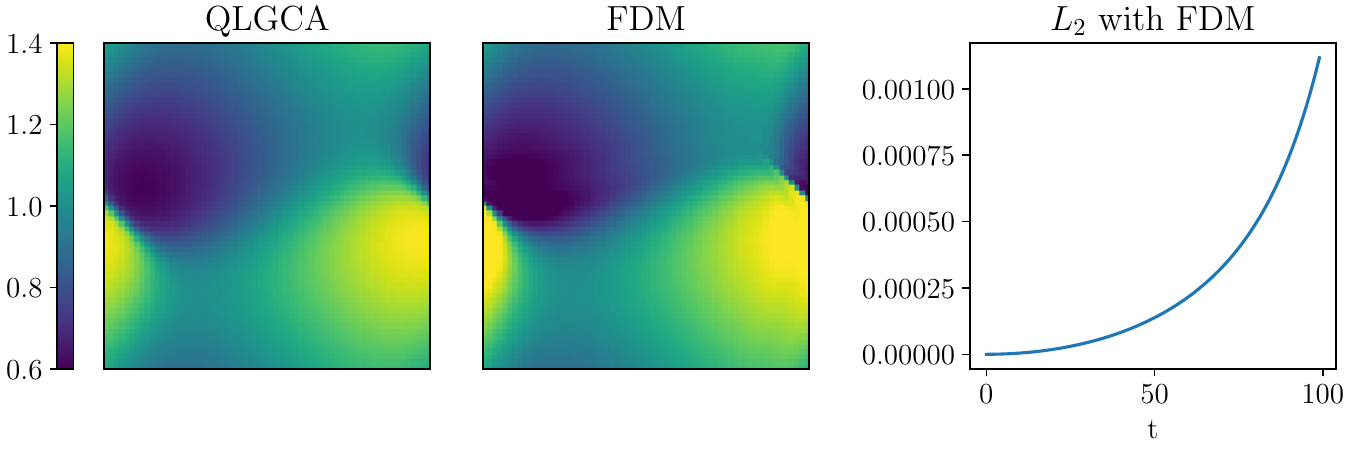}
    \caption{Comaprison between QLG simulations and FDM solving Eq. \ref{eq:final_pde_2d} for different velcoty sets. Simulations were made with a 64$\times$64 square grid for different discrete velocity sets initialized as in Eq. \ref{eq:initial_cos}. The blue (red) arrow represents the streaming direction of $f_0$ ($f_1$).}
    \label{fig:2D_simulations_withFDM_2}
\end{figure*}

\end{document}